\documentclass[a4paper,11pt]{article}
\pdfoutput=1
\usepackage{jheppub}
\usepackage[T1]{fontenc}
\usepackage{multirow}
\usepackage{booktabs}
\usepackage{mathtools}
\usepackage{adjustbox}
\usepackage{array}
\newcommand{\eqal}[1]{\begin{align}#1\end{align}}

\begin{document}

\title{\boldmath Neutrino non-standard interactions meet precision measurements of $N_{\rm eff}$}

\author[a]{Yong Du,}
\author[a,b,c,d,e,1]{Jiang-Hao Yu\note{Corresponding author.}}
\affiliation[a]{CAS Key Laboratory of Theoretical Physics, Institute of Theoretical Physics, Chinese Academy of Sciences, Beijing 100190, P. R. China}
\affiliation[b]{School of Physical Sciences, University of Chinese Academy of Sciences, Beijing 100049, P.R. China}
\affiliation[c]{Center for High Energy Physics, Peking University, Beijing 100871, China}
\affiliation[d]{School of Fundamental Physics and Mathematical Sciences, Hangzhou Institute for Advanced Study, UCAS, Hangzhou 310024, China}
\affiliation[e]{International Centre for Theoretical Physics Asia-Pacific, Beijing/Hangzhou, China}

\emailAdd{yongdu@itp.ac.cn}
\emailAdd{jhyu@itp.ac.cn}

\abstract{The number of relativistic species, $N_{\rm eff}$, has been precisely calculated in the standard model, and would be measured to the percent level by CMB-S4 in future. Neutral-current non-standard interactions would affect neutrino decoupling in the early Universe, thus modifying $N_{\rm eff}$. We parameterize those operators up to dimension-7 in the effective field theory framework, and then provide a complete, generic and analytical dictionary for the collision term integrals. From precision measurements of $N_{\rm eff}$, the most stringent constraint is obtained for the dimension-6 vector-type neutrino-electron operator, whose scale is constrained to be above about 195 (331)\,GeV from Planck (CMB-S4). We find our results complementary to other experiments like neutrino coherent scattering, neutrino oscillation, collider, and neutrino deep inelastic scattering experiments.
}

\arxivnumber{2101.10475}
\maketitle
\flushbottom

\section{Introduction}\label{sec:intro}
The great triumph of the Standard Model (SM) of particle physics was the discovery of the Higgs particle in 2012\,\cite{Aad:2012tfa,Chatrchyan:2012ufa}. However, SM can not be the complete theory as there are still several unsolved puzzles, such as neutrino masses, dark matter and baryon asymmetry of the Universe, which require new physics beyond the SM. Tremendous new physics models have been invented and studied to address these issues, yet no definite signals of any of these models have been observed at colliders or from low-energy precision measurements. This has in turn motivated physicists to search for new physics in a model-independent and systematic way. 

Effective Field Theories (EFTs) provide such a systematic and model-independent framework for the study of new physics, especially if its characteristic scale is above the weak scale. The EFT, obtained by integrating out the newly introduced heavy particles to the SM, is called the SM EFT (SMEFT)\,\cite{Weinberg:1979sa,Buchmuller:1985jz,Grzadkowski:2010es,Lehman:2014jma,Li:2020gnx,Murphy:2020rsh,Li:2020xlh,Liao:2020jmn,Liao:2016hru}, which respects the SM gauge group and is valid until down to the weak scale. Below the weak scale, the corresponding EFT is the Low-energy EFT (LEFT)\,\cite{Jenkins:2017jig,Liao:2020zyx,Li:2020tsi,Murphy:2020cly}, where the top quark, the SU(2) gauge bosons and the Higgs particle of the SMEFT are all integrated out. As a consequence, the Lagrangian of the LEFT respects the $\rm SU(3)_c\times U(1)_{\rm EM}$ gauge group. 

Since the discovery of neutrino oscillations\,\cite{Davis:1968cp,Ahmad:2001an,Fukuda:1998mi,An:2012eh,Ahn:2002up,Michael:2006rx}, neutrino non-standard interactions (NSIs), firstly discussed in Refs.\,\cite{Wolfenstein:1977ue,Mikheev:1986gs} and {\color{black}nicely} reviewed in Refs.\,\cite{Davidson:2003ha,Ohlsson:2012kf,Farzan:2017xzy,Dev:2019anc,Abazajian:2012ys}, have gained significant attention in recent years and can be described by the LEFT framework. Very stringent constraints on these NSI operators have been obtained, see, for example, Refs.\,\cite{Agrawal:2013hya,Nelson:2013pqa,Pobbe:2017wrj,Choudhury:2018xsm,Friedland:2011za,Babu:2020nna,Falkowski:2017pss,Escrihuela:2011cf,Coloma:2017ncl,Altmannshofer:2018xyo,Babu:2019mfe,Khan:2019cvi,Papoulias:2019xaw,Canas:2019fjw,Falkowski:2019xoe} for recent theoretical studies and Refs.\,\cite{Abbiendi:2003dh,Breitweg:1999ssa,Adloff:2000dp,Khachatryan:2014rra,Aad:2015zva} for experimental investigation. On the other hand, since these NSI operators can be matched to SMEFT operators, constraints on the NSIs from low-energy experiments can also be translated into constraints on the SMEFT operators, thus also on the UV models. While we are not interested in UV completion of neutrino NSIs in this work, we comment that these NSI operators can be induced, for example, from the leptoquark model\,\cite{Dorsner:2016wpm} and/or the $\rm U(1)'$ models, see, for example, the discussion in Ref.\,\cite{Wise:2014oea}.

These neutrino NSI operators can be generically classified into charge-current (CC) and neutral-current (NC) ones.\footnote{In the case of generic neutrino interactions, see Refs.\,\cite{Lindner:2016wff,Rodejohann:2017vup,Bischer:2018zcz,Bischer:2019ttk,Khan:2019jvr}.} In Ref.\,\cite{Biggio:2009nt}, bounds on the CC NSIs were obtained from the Cabibbo–Kobayashi–Maskawa (CKM)\,\cite{Cabibbo:1963yz,Kobayashi:1973fv} unitarity, weak universality tests from pion decay\,\cite{Loinaz:2004qc}, short-baseline neutrino oscillation experiments KARMEN\,\cite{Eitel:2000by} and NOMAD\,\cite{Astier:2001yj,Astier:2003gs}, and loop corrections to $\mu\to e$ conversion in gold\,\cite{Zyla:2020zbs}. Very recently, CC NSIs were studied in Ref.\,\cite{Terol-Calvo:2019vck,Du:2020dwr} within the SMEFT framework. 

For NC NSIs at dimension-6, one-loop electroweak radiative corrections was recently calculated in Ref.\,\cite{Hill:2019xqk} within the SM, including two-loop matching and three-loop running for the lepton sector. Constraints from collider searches, dark matter direct detection experiments, and superbeam experiments can be found in Refs.\,\cite{Harnik:2012ni,Cadeddu:2018izq,Huang:2018nxj,Shoemaker:2018vii,AristizabalSierra:2017joc,Gonzalez-Garcia:2018dep,Dutta:2017nht,Bertuzzo:2017tuf,Dent:2016wcr,Cerdeno:2016sfi,Coloma:2014hka,Pospelov:2013rha,Pospelov:2012gm,Kopp:2007ne,Liu:2020emq}. See Refs.\,\cite{Falkowski:2018dmy,Bischer:2018zcz,Pandey:2019apj,Deepthi:2016erc,Deepthi:2017gxg} for neutrino trident production and neutrino-electron scattering from DUNE, Refs.\,\cite{Davidson:2003ha,Barranco:2005ps,Barranco:2007ej,Bolanos:2008km,Biggio:2009nt,Lei:2019nma,Esmaili:2013fva,Friedland:2004ah,Friedland:2005vy,Khan:2017oxw} for oscillation experiments, Ref.\,\cite{Davidson:2003ha,Biggio:2009kv} for loop bounds on dimension-6 electron-neutrino contact operators, Refs.\,\cite{Tomalak:2020zfh,Denton:2020hop,Hoferichter:2020osn,Akimov:2017ade,Altmannshofer:2018xyo,Miranda:2019skf,Deniz:2010mp,Khan:2016uon} for neutrino coherent scattering experiments, and Ref.\,\cite{Ismail:2020yqc} for FASER$\nu$. Note also that the dimension-6 NC electron self-interacting NSIs could modify the weak mixing angle. This angle would be very precisely measured by the upcoming low-energy MOLLER experiment at the Jefferson Lab\,\cite{Benesch:2014bas} and the planned P2 experiment at MESA\,\cite{Berger:2015aaa}. Recently, SM prediction of the weak mixing angle at two-loop has been obtained in Ref.\,\cite{Du:2019hwm}. For NC neutrino NSIs up to dimension-7, part of them was previously investigated in Refs.\,\cite{Esteban:2018ppq,Farzan:2018gtr,Billard:2018jnl,AristizabalSierra:2018eqm,Kosmas:2017tsq,Dent:2017mpr,Liao:2017uzy,Dent:2016wcr,Lindner:2016wff}, while a relatively more comprehensive study was recently presented in Ref.\,\cite{Altmannshofer:2018xyo}.

However, not all NC NSI operators up to dimension-7 are bounded from Ref.\,\cite{Altmannshofer:2018xyo} or existing work. For example, dimension-6 neutrino self-interacting operators are not studied, since previous work mainly focuses on neutrino oscillation, neutrino coherent scattering and collider experiments etc., which are insensitive to these operators. Furthermore, at dimension-7, only neutrino-photon, neutrino-gluon, and neutrino-quark operators are investigated\,\cite{Altmannshofer:2018xyo}, while the dimension-7 neutrino-electron operators are not yet considered to be constrained.

In the early Universe where only neutrinos, electrons, positrons, and photons are present, these neutrino-neutrino, neutrino-electron/positron and neutrino-photon NC NSIs would affect neutrino decoupling, thus modifying the effective number of relativistic degrees of freedom, viz., $N_{\rm eff}$. In light of the precision measurements of $N_{\rm eff}$ from LEP\,\cite{ALEPH:2005ab} and Planck\,\cite{Aghanim:2018eyx}, the upcoming SPT-3G\,\cite{Benson:2014qhw} and the Simon Observatory\,\cite{Ade:2018sbj}, as well as the proposal from Cosmic Microwave Background-Stage 4 (CMB-S4)\,\cite{Abazajian:2016yjj}, CORE\,\cite{DiValentino:2016foa}, PICO\,\cite{Hanany:2019lle} and CMB-HD\,\cite{Sehgal:2019ewc}, one naturally expects constraints on these NC NSIs from $N_{\rm eff}$. 

In this work, we investigate all kinds of neutrino-neutrino, neutrino-electron/positron and neutrino-photon NC NSI operators up to dimension-7, as well as their impact on $N_{\rm eff}$. Since the light mediator directly serves as one additional degree of freedom and thus resulting in large $N_{\rm eff}$, these NC NSI operators we consider in this work are assumed to be induced by integrating out some heavy new physics above $\sim\mathcal{O}(\rm100\,MeV)$ that is about the muon mass or heavier.

To obtain new physics corrections to $N_{\rm eff}$, the SM prediction to $N_{\rm eff}$ has to be known precisely in the first place. However, it has been known for a long time that the precision calculation of $N_{\rm eff}$ is very challenging. Within the SM, the precision calculation of $N_{\rm eff}$ has been carried out through the density matrix formalism. Due to its complexity, however, the density matrix formalism is very difficult to generalize to other scenarios, for example, when new physics is present. For recent development of precision calculation of $N_{\rm eff}$, see Refs.\,\cite{Bennett:2019ewm,Bennett:2020zkv,Akita:2020szl,Escudero:2018mvt,Escudero:2020dfa}.

In this work, we adopt the strategy developed in Refs.\,\cite{Escudero:2018mvt,Escudero:2020dfa} that reproduces the SM prediction of $N_{\rm eff}$, works fast, and can be easily generalized to include effects from various new physics. To find corrections to $N_{\rm eff}$ from some new physics, this strategy has already been applied in Refs.\,\cite{Luo:2020sho,Luo:2020fdt} with the introduction of right-handed partners of neutrinos, Ref.\,\cite{Kelly:2020aks,Adshead:2020ekg} with dark matter and/or sterile neutrinos, Ref.\,\cite{Li:2020roy} for dark photon, Ref.\,\cite{Venzor:2020ova} with the introduction of neutrino-scalar interactions, Ref.\,\cite{Froustey:2020mcq} with the inclusion of neutrino flavor oscillation and primordial nucleosynthesis, and Ref.\,\cite{Ibe:2020dly} with a light $Z'$ to explain the recent XENON1T excess\,\cite{Aprile:2020tmw}. Applying this strategy to the calculation of $N_{\rm eff}$ with the inclusion of NC NSIs up to dimension-7, in this work, we
\begin{itemize}
\item provide a complete, generic and analytical dictionary for the collision term integrals in section\,\ref{sec:SMEFT}. This dictionary can be used directly for computing corrections to $N_{\rm eff}$ from some new physics, either in the EFT framework up to dimension-7, or in some UV models as long as the new physics is above $\sim\mathcal{O}(\rm100\,MeV)$;
\item present our constraints on the NC neutrino NSI operators up to dimension-7 in section\,\ref{sec:EFTConstraints}, and also compare our results with previous ones.
\end{itemize}

The rest of this work is organized as follows. We briefly review neutrino decoupling in the early Universe and the definition of $N_{\rm eff}$ in section\,\ref{sec:revneff}. In section\,\ref{setup:BoltzEQ}, we discuss the strategy developed in Refs.\,\cite{Escudero:2018mvt,Escudero:2020dfa}, and then summarize our strategy for calculating the collision terms integrals. Since these collision term integrals are essential to boost the calculation of $N_{\rm eff}$, we provide a complete generic and analytical dictionary of the collision term integrals, as well as the NSI operators we study in this work in section\,\ref{sec:SMEFT}. Constraints on these NC NSI operators are presented in section\,\ref{sec:EFTConstraints}. We conclude in section\,\ref{sec:con}.

\section{Brief review of neutrino decoupling and $N_{\rm {eff}}$}\label{sec:revneff}
In the early Universe when the temperature is above $\mathcal{O}(10)$\,MeV and below the muon mass, electrons, positrons, neutrinos and photons are in thermal equilibrium from electroweak interactions. As the Universe expands and the temperature cools down, neutrinos decouple from the rest of the plasma at around $T_{\rm dec}=2$\,MeV. The neutrinos then undergo simple dilution from the expansion of the Universe, while $e^\pm$ and photons are still in thermal equilibrium. However, when the photon temperature cools further down below the electron mass $m_e$, $\gamma \gamma \to e^+ e^-$ becomes suppressed while the inverse process is still permitted, heating up the photons. 

The number of relativistic degrees of freedom during this period can be parameterized by $N_{\rm eff}$\,\cite{Shvartsman:1969mm,Steigman:1977kc,Mangano:2001iu}:
\eqal{
\rho_R=\left[ 1+\frac{7}{8}\left(\frac{4}{11}\right)^{\frac{4}{3}} N_{\rm eff}\right]\rho_\gamma
}
with $\rho_\gamma$ the photon energy density, and $\rho_R$ the total energy density from all relativistic species during this epoch. Equivalently,
\eqal{
N_{\rm eff} \equiv\left(\frac{\rho_{R}-\rho_{\gamma}}{\rho_{\nu}^{0}}\right)\left(\frac{\rho_{\gamma}^{0}}{\rho_{\gamma}}\right),\label{eq:neff}
}
with $\rho_{\nu}^0$ the energy density of a single massless neutrino, and $\rho_\gamma^0$ the energy density of photons in the instantaneous decoupling limit. Obviously, in the instantaneous limit, $\rho_\gamma^0=\rho_\gamma$ and $\rho_R=3\rho_\nu^0+\rho^\gamma$, resulting in the well-known $N_{\rm eff}=3$.

On the other hand, due to the tininess of neutrino masses, the three flavor neutrinos can be effectively taken as massless, permitting to express $N_{\rm eff}$ in eq.\,\eqref{eq:neff} also in terms of the photon temperature $T_{\gamma}$ and the neutrino temperature $T_{\nu}$ as, upon assuming $T_\gamma=T_e$ which is valid since photons and electrons are tightly coupled during neutrino decoupling,
\eqal{
N_{\rm eff}=3\left(\frac{11}{4}\right)^{4 / 3}\left(\frac{T_{\nu}}{T_{\gamma}}\right)^{4}.\label{eq:nefffinal}
}
Similarly, in the instantaneous decoupling limit, ${T_{\nu}}/{T_{\gamma}}=(4/11)^{1/3}$\,\cite{Kolb:1990vq} and once again $N_{\rm eff}=3$.

However, it has been known for decades that the instantaneous decoupling picture is not accurate. Indeed, neutrinos are still slightly interacting with the electromagnetic plasma, and neutrino oscillations are also active during neutrino decoupling\,\cite{deSalas:2016ztq,Mangano:2005cc,Hannestad:2001iy,Dolgov:2002ab}. Furthermore, the electromagnetic plasma also receives corrections from finite temperature QED corrections. Taking all these effects into account, one finds $N_{\rm eff}=3.044$\,\cite{Akita:2020szl,Froustey:2020mcq}. Corrections from these effects will be discussed further in detail in sections\,\ref{sec:exeff1} and \ref{sec:meandSpinCorrections}.

\section{Setup of the Boltzmann equation}\label{setup:BoltzEQ}
Evolution of phase space distribution (PSD) of any particle in the early Universe is governed by the Boltzmann equation, which we briefly review in this subsection. As mentioned in the introduction, we follow the discussion in Ref.\,\cite{Escudero:2020dfa}, which simplifies the calculation of $N_{\rm eff}$ significantly and reproduces the prediction for $N_{\rm eff}$ by using the density matrix formalism.

\subsection{The Boltzmann equation}
The Boltzmann equation reads
\eqal{\frac{\partial f_i}{\partial t}-H p \frac{\partial f_i}{\partial p}=\mathcal{C}[f_i],\label{eq:boltzmannL}}
with $f_i(p,t)$ the PSD for particle $i$, $H$ the Planck constant that accounts for the dilution effect from the expansion of the Universe, and $\mathcal{C}$ the collision term defined as\footnote{Note that in our setup, we include the symmetry factor in the definition of $\langle\mathcal{M}^{2}\rangle$ throughout this work.}
\eqal{
\mathcal{C}\left[f_{i}\right] \equiv & \frac{1}{2 E_{i}} \sum_{X, Y} \int \prod_{i,j} d \Pi_{X_{i}} d \Pi_{Y_{j}}(2 \pi)^{4} \delta^{4}\left(p_{i}+p_{X}-p_{Y}\right) \nonumber\\
& \times\left( \langle\mathcal{M}^{2}\rangle_{Y \rightarrow i+X} \prod_{i,j} f_{Y_{j}}\left[1 \pm f_{i}\right] \left[1 \pm f_{X_{i}}\right] -  \langle\mathcal{M}^{2}\rangle_{i+X \rightarrow Y}  \prod_{i,j} f_{i} f_{X_{i}}\left[1 \pm f_{Y_{j}}\right]\right),\label{collterm}
}
where $d \Pi_{{i}}\equiv d^3p_{i}/[(2\pi)^3 2E_{i}]$ and ``+ ($-$)'' is for bosonic (fermionic) particles. Note that the difference in the last line above correctly accounts for the production and annihilation of particle $i$.

Upon integrating over the phase space of particle $i$ on both sides of eq.\,\eqref{eq:boltzmannL}, one finds\footnote{Without any ambiguity, we suppress the index $i$ starting from here.}
\eqal{
\frac{d n}{d t}+3 H n &=\frac{\delta n}{\delta t}\equiv\int g\frac{d^{3} p}{(2 \pi)^{3}} \mathcal{C}[f],\label{numdensity} \\
\frac{d \rho}{d t}+3 H(\rho+p) &=\frac{\delta \rho}{\delta t}\equiv\int g E \frac{d^{3} p}{(2 \pi)^{3}} \mathcal{C}[f],\label{energydensity}
}
where $g$ is the intrinsic degree of freedom of particle $i$, $E$ is its energy, and $n$ and $\rho$ are the number and the energy densities of particle $i$ respectively. Note that after the phase space integration on the right hand side of eqs.(\ref{numdensity}-\ref{energydensity}), ${\delta n}/{\delta t}$ and ${\delta \rho}/{\delta t}$ are functions of the temperature $T$, the chemical potential $\mu$ and the model parameters only.\footnote{With the inclusion of NSI operators, it will also depend on the scale of new physics and the Wilson coefficients.} Thus, in terms of the Hubble parameter, one can readily obtain the following equations through the application of the chain rule:
\eqal{
\frac{d T}{d t}=&\frac{1}{\frac{\partial n}{\partial \mu} \frac{\partial \rho}{\partial T}-\frac{\partial n}{\partial T} \frac{\partial \rho}{\partial \mu}}\left[-3 H\left((p+\rho) \frac{\partial n}{\partial \mu}-n \frac{\partial \rho}{\partial \mu}\right)+\frac{\partial n}{\partial \mu} \frac{\delta \rho}{\delta t}-\frac{\partial \rho}{\partial \mu} \frac{\delta n}{\delta t}\right]\label{Temevolv} \\
\frac{d \mu}{d t}=&\frac{-1}{\frac{\partial n}{\partial \mu} \frac{\partial \rho}{\partial T}-\frac{\partial n}{\partial T} \frac{\partial \rho}{\partial \mu}}\left[-3 H\left((p+\rho) \frac{\partial n}{\partial T}-n \frac{\partial \rho}{\partial T}\right)+\frac{\partial n}{\partial T} \frac{\delta \rho}{\delta t}-\frac{\partial \rho}{\partial T} \frac{\delta n}{\delta t}\right]\label{chemevolv}.
}
These two equations effectively describe the evolution of $T$ and $\mu$ for any particles in the early Universe, and can thus be used to solve the decoupling of neutrinos from the rest of the plasma as we will see later in this section.

\subsubsection{Evolution of $T_\gamma$, $T_\nu$ and $\mu_\nu$ in the SM}\label{sec:exeff1}
At the time of neutrino decoupling, since photons and electrons are still tightly coupled, one can safely set $\mu_\gamma=\mu_e=0$ and $T_\gamma = T_e$. By applying eqs.\,(\ref{Temevolv}-\ref{chemevolv}), one obtains\,\cite{Escudero:2020dfa}
\eqal{
\frac{d T_{\gamma}}{d t}=&-\frac{4 H \rho_{\gamma}+3 H\left(\rho_{e}+p_{e}\right)+\frac{\delta \rho_{\nu e}}{\delta t} + \frac{\delta \rho_{\nu \mu}}{\delta t} + \frac{\delta \rho_{\nu \tau}}{\delta t} }{\frac{\partial \rho_{\gamma}}{\partial T_{\gamma}}+\frac{\partial \rho_{e}}{\partial T_{\gamma}}},\\
\frac{d T_{\nu_{\alpha}}}{d t}=&-H T_{\nu_{\alpha}}+\frac{\delta \rho_{\nu_{\alpha}}}{\delta t} / \frac{\partial \rho_{\nu_{\alpha}}}{\partial T_{\nu_{\alpha}}}, \quad \quad \alpha = e,\mu,\tau.\label{TvFinal}
}

Note that the above equation for $T_\gamma$ is derived assuming the finite temperature corrections are negligible, while it has been known for a long time that this is not the case especially given the precision measurements of $N_{\rm eff}$ from future experiments. To be clearer, in the future, $N_{\rm eff}$ will be measured to the percent level, while finite temperature corrections to $N_{\rm eff}$ is also at the percent level\,\cite{Abazajian:2016yjj,Abazajian:2019tiv,Abitbol:2017nao,Abazajian:2013oma,DiValentino:2016foa,Hanany:2019lle, Sehgal:2019ewc, Abazajian:2019eic}. Therefore, to correctly interpret the results from future experiments and/or to disentangle contributions to $N_{\rm eff}$ from any potential new physics from the SM, the QED corrections have to be included.

The leading-order QED corrections were obtained decades ago\,\cite{Heckler:1994tv,Fornengo:1997wa}, and higher-order corrections up to $\mathcal{O}(e^4)$ were recently calculated in Ref.\,\cite{Bennett:2019ewm}, where the authors found corrections to $N_{\rm eff}$ are about $-$0.0009 and $10^{-6}$ at $\mathcal{O}(e^3)$ and $\mathcal{O}(e^4)$ respectively. Since both corrections at $\mathcal{O}(e^3)$ and $\mathcal{O}(e^4)$ exceed the proposed precision target of the future experiments, we neglect those in our setup and only keep the finite temperature corrections up to $\mathcal{O}(e^2)$. On the other hand, neutrino oscillations also lead to a correction to $N_{\rm eff}$, which is about 0.0007 as reported in Ref.\,\cite{Mangano:2005cc,deSalas:2016ztq,Gariazzo:2019gyi}. Note that, as was pointed out in Ref.\,\cite{Bennett:2019ewm}, since contributions to $N_{\rm eff}$ from neutrino oscillations and the finite temperature corrections at $\mathcal{O}(e^3)$ are comparable, they shall both be included for a consistent precision calculation of $N_{\rm eff}$. However, as stated above, due to their smallness, we also neglect contributions from neutrino oscillations in this work.

To conclude this subsection, we show the result for $T_\gamma$ with the inclusion of the aforementioned finite temperature corrections following the notations of Refs.\,\cite{Bennett:2019ewm,Escudero:2020dfa}:
\eqal{
\frac{d T_{\gamma}}{d t}=-\frac{4 H \rho_{\gamma}+3 H\left(\rho_{e}+p_{e}\right)+3 H T_{\gamma} \frac{d P_{\text {int }}}{d T_{\gamma}}+\frac{\delta \rho_{\nu e}}{\delta t} + \frac{\delta \rho_{\nu \mu}}{\delta t} + \frac{\delta \rho_{\nu \tau}}{\delta t}}{\frac{\partial \rho_{\gamma}}{\partial T_{\gamma}}+\frac{\partial \rho_{e}}{\partial T_{\gamma}}+T_{\gamma} \frac{d^{2} P_{\text {int }}}{d T_{\gamma}^{2}}},\label{TgammaFinal}
}
where $P_{\rm int}$ and $\rho_{\rm int}\equiv -P_{\rm int} + dP_{\rm int}/d\ln T_\gamma$ are finite temperature corrections to the electromagnetic pressure and the electromagnetic energy density respectively, whose analytical expressions can be found in Ref.\,\cite{Bennett:2019ewm}.

\subsection{Brief review of the collision term integrals}\label{sec:PSIntegral}
From eqs.\eqref{TgammaFinal} and \eqref{TvFinal}, one can then solve $T_\gamma (t)$ and $T_\nu (t)$, and thus $N_{\rm eff}$ at the time of neutrino decoupling. From eqs.(\ref{numdensity}-\ref{energydensity}), we conclude that to solve $T_\gamma (t)$ and $T_\nu (t)$, the remaining task is to first finish these phase space integrals, which are in general very challenging with no analytical expressions. This in turn slows down numerical calculation of $N_{\rm eff}$, especially in the presence of new physics. However, as pointed out in Ref.\,\cite{Escudero:2020dfa}, analytical results for those collision term integrals exist in the Maxwell-Boltzmann limit, as a result, numerical calculation of $N_{\rm eff}$ can be boosted significantly. For specific processes in the SM, the author of Refs.\,\cite{Escudero:2020dfa} presented analytical results for the collision terms integrals in Ref.\,\cite{Escudero:2018mvt}, which, however, can not be generalized to processes in the presence of new physics. In light of this, and since the analytical forms of the collision term integrals are essential to boost the numerical calculation of $N_{\rm eff}$, we present a full generic and analytical dictionary for the collision term integrals in section\,\ref{sec:SMEFT}, while present the method we use to obtain these collision term integrals in this subsection.

To start, we consider the collision term integrals for $2\to2$ processes as these are the only interaction types relevant for $N_{\rm eff}$ calculation in the SM and with the inclusion of NSI operators considered in this work. \footnote{For decay or inverse decay, the collision term integral, defined is eq.\,\eqref{eq:colltermintdef} is a nine-fold one that can be reduced to a two-fold integral. There are many literatures in the past discussing the collision term integral, see, for example, Refs.\cite{Hannestad:1995rs,Dolgov:1997mb,Dolgov:1998sf,Birrell:2014uka,Oldengott:2017fhy,Oldengott:2014qra,Grohs:2015tfy,Bennett:2019ewm,Yunis:2020woq,Kreisch:2019yzn,Mangano:2005cc, deSalas:2016ztq, Gariazzo:2019gyi,Esposito:2000hi, Mangano:2001iu,Froustey:2019owm}.} Upon leaving out the irrelevant factor $g$, for a generic process $1 + 2 \to 3 + 4$, one can write the collision term integrals on the right hand of eqs.(\ref{numdensity}-\ref{energydensity}) generically as
\eqal{
C^{(j)}\equiv\int E^j_1\frac{d^{3} p_1}{(2 \pi)^{3}} \mathcal{C}[f_1]\quad \Leftrightarrow \quad \left\{\begin{array}{cl}j=0 & \text{, for number density} \\ j=1 & \text{, for energy density} \end{array}\right.,\label{eq:colltermintdef}
}
with $p_i$ and $E_i$ the four-momentum and the energy of the $i$-th particle. To simplify the collision term integral, we reproduce the results presented in Appendix D of Ref.\,\cite{Fradette:2018hhl} and cite the result here:\footnote{We assume CP conservation that allows the factorization of $\langle\mathcal{M}^{2}\rangle_{1+2\to3+4}$, which is a well-justified approximation for our purpose here.}

\eqal{
C^{(j)} = & \frac{1}{2(2 \pi)^{6}} \int E_1^jd E_{1} d E_{2} d E_{3} \cdot \left(|\vec{p}_{1}| \,|\vec{p}_{2}|\, |\vec{p}_{3}|\right)\cdot\Theta\left(Q+|\vec{p}_{1}|^{2}+|\vec{p}_{2}|^{2}+|\vec{p}_{3}|^{2}+2 \gamma\right)\nonumber\\
&\quad\quad\times \left(\int_{\max \left(-1, \cos \theta_{-}\right)}^{\min \left(1, \cos \theta_{+}\right)} d(\cos \theta) \int_{\cos \alpha_{-}}^{\cos \alpha_{+}} d(\cos \alpha) \frac{\langle\mathcal{M}^{2}\rangle_{1+2\to3+4}}{\sqrt{a \cos ^{2} \alpha+b \cos \alpha+c}}\right.\nonumber\\
&\quad\quad\quad\quad\left.\left. \times \left[f_3f_4(1\pm f_1)(1\pm f_2) - f_1 f_2 (1\pm f_3)(1\pm f_4)\right]{\frac{}{}}\right)\right|_{p_4\to p_1+p_2-p_3},\label{CollIntFinal}
}
where $\alpha$ ($\theta$) is the angle between $\vec{p}_1$ and $\vec{p}_2$ ($\vec{p}_3$), and we define
\eqal{
Q\equiv& \,m_1^2+m_2^2+m_3^2-m_4^2,\\
\gamma \equiv&\, E_1E_2 - E_1 E_3 - E_2 E_3,\\
\omega\equiv&\, Q + 2\gamma + 2|\vec{p}_{1}||\vec{p}_{1}|\cos\theta,\\
a\equiv&\,-4 |\vec{p}_{2}|^{2}\left(|\vec{p}_{1}|^{2}+|\vec{p}_{3}|^{2}-2 |\vec{p}_{1}| |\vec{p}_{3}| \cos \theta\right), \\
b\equiv&\,4 |\vec{p}_{2}|\left(|\vec{p}_{1}|-|\vec{p}_{3}| \cos \theta\right) \omega, \\
c\equiv&\,4 |\vec{p}_{2}|^{2} |\vec{p}_{3}|^{2} \sin ^{2} \theta-\omega^{2}.
}
Note that $a\le0$ from above definition, and the integrating regions for $\alpha$ and $\theta$ are altered, where
\eqal{
\cos\alpha_{\pm}=&\,\frac{-b \mp \sqrt{b^{2}-4 a c}}{2 a},\\
\cos\theta_{\pm}=&\,-\frac{Q+2 |\vec{p}_{2}|^{2}+2 \gamma \mp 2 |\vec{p}_{2}| \sqrt{Q+|\vec{p}_{1}|^{2}+|\vec{p}_{2}|^{2}+|\vec{p}_{3}|^{2}+2 \gamma}}{2 |\vec{p}_{1}| |\vec{p}_{3}|},
}
resulting from the requirement of the existence of physical solutions to the collision term integral. Note also that $\langle\mathcal{M}^{2}\rangle_{1+2\to3+4}$ is in general a function of $p_{ij}$, defined as
\eqal{p_{ij}\equiv p_i\cdot p_j,\quad (i,j=1,\dots,4),\label{def:pij}}
which also depends on the angles $\alpha$ and $\theta$, making the integral in eq.\,\eqref{CollIntFinal} too cumbersome to be completed.

Surprisingly, one huge simplification that eventually allows the completion of the integral in eq.\,\eqref{CollIntFinal} can be realized when (1) all the particles involved are massless, i.e., $m_i=0$ ($i=1,\dots,4$), leading to
\eqal{
Q=0, \quad {\rm min}(1,\cos\theta_+) = 1 = \cos\theta_+,
}
and (2) all the particles obey the Maxwell-Boltzmann distribution, permitting analytical expressions for almost all possible forms of $\langle\mathcal{M}^{2}\rangle_{1+2\to3+4}$. The only exception is when there exists a light mediator in the $t$ and/or the $u$ channels, where it has been well-known that IR divergence emerges when all external particles become massless, the Compton scattering for example.\footnote{Note, however, that even in the case with a light mediator in the $s$ channel, there is no IR divergence and analytical results for the collision term integrals can always be obtained.} However, this IR divergence has to cancel out for any sufficiently inclusive quantities, as is guaranteed by the Kinoshita-Lee-Nauenberg (KLN) theorem\,\cite{Kinoshita:1962ur,Lee:1964is}. Recently, it is also shown in Ref.\,\cite{Frye:2018xjj} that to have IR finiteness, one does not necessarily need to sum over both the initial and the final states as stated by the KLN theorem, rather, one only needs to sum over all possible final (initial) state for a given fixed initial (final) state, as long as the forward scattering is also included.

In this work, since we are interested in constraints on new physics from $N_{\rm eff}$ in a model independent manner within the EFT framework, we will mainly focus on scenarios with heavy mediators such that the IR divergence issue mentioned above never shows up. The only exception is the dimension-5 neutrino magnetic dipole operator, which we will discuss in section\,\ref{sec:SMEFT}. Furthermore, in order to obtain analytical results, as discussed in last paragraph, we assume all particles (1) are massless, and (2) obey the Maxwell-Boltzmann distribution only when calculating the collision term integrals from eq.\,\eqref{CollIntFinal}. We then present a complete dictionary for all possible $\langle\mathcal{M}^{2}\rangle_{1+2\to3+4}$ up to products with three $p_{ij}$ in $\langle\mathcal{M}^{2}\rangle_{1+2\to3+4}$ in section\,\ref{sec:SMEFT}, {\color{black}which are also provided in auxiliary {\tt Mathematica} notebook files}. Corrections from non-vanishing masses and Fermi-Dirac/Bose-Einstein distribution are also discussed in section\,\ref{sec:SMEFT}.

\section{EFT operators and the collision term integrals}\label{sec:SMEFT}
As no new particles have been observed after the discovery of the Higgs particle in 2012\,\cite{Aad:2012tfa,Chatrchyan:2012xdj}, EFTs have become the natural framework for the study of any new heavy physics. In this work, we are interested in corrections to $N_{\rm eff}$ from higher-dimensional operators in the early Universe. The active degrees of freedom at that time are neutrinos, photons, electrons and positrons. Since the neutrinos decouple from the rest of the plasma at around 2\,MeV, for the EFTs to be valid, the potential new physics could be as light as $\sim\mathcal{O}(100\,\rm MeV)$ that is about the muon mass. Note that the lower bound of the new physics scale would also be constrained from, for example, Big Bang Nucleosynthesis. Given that future experiments like CMB-S4 could constrain $\Delta N_{\rm eff}<0.06$ at 95\% CL\,\cite{Abazajian:2016yjj,Abazajian:2019tiv,Abitbol:2017nao,Abazajian:2019eic} with $\Delta N_{\rm eff}$ the corrections to SM prediction of $N_{\rm eff}$, one naturally expects the EFT operators could also be constrained from the precision measurements of $N_{\rm eff}$.

In this section, we will first enumerate the relevant EFT operators up to dimension-7 in section\,\ref{sec:EFTOperators}. The resulting invariant amplitudes from these operators turn out to be functions of $p_{ij}$ defined in eq.\,\eqref{def:pij}, model parameters and the Wilson coefficients. Depending on how explicitly the $\langle \mathcal{M}^2\rangle$ depends on $p_{ij}$, the collision term integral in eq.\,\eqref{CollIntFinal} needs to be calculated case by case. From momentum-energy conservation, the redundancy in collision term integral computation can be reduced to a set of limited number of bases as presented in section\,\ref{sec:EFTOpeBas}. Starting from these bases, we then present a complete dictionary of the collision term integrals in section\,\ref{sec:ComDictColl}.

\subsection{List of relevant EFT operators}\label{sec:EFTOperators}

\begin{table}
 \renewcommand{\arraystretch}{1.5}
\centering
\begin{tabular}{|c|l|c|}
\hline
Dimensions & Operators & Wilson coefficients\\
\hline
{\rm dimension-5} & $\mathcal{O}_1^{(5)}=\frac{e}{8 \pi^{2}}\left(\bar{\nu}_{\beta} \sigma^{\mu \nu} P_{L} \nu_{\alpha}\right) F_{\mu \nu}$ & $C_1^{(5)}$ \\
\hline
\multirow{5}{*}{\rm dimension-6} & $\mathcal{O}_{1,f}^{(6)}=\left(\bar{\nu}_{\beta} \gamma_{\mu} P_{L} \nu_{\alpha}\right)\left(\bar{f} \gamma^{\mu} f\right)$ & $C_{1,f}^{(6)}$ \\
& $\mathcal{O}_{2,f}^{(6)}=\left(\bar{\nu}_{\beta} \gamma_{\mu} P_{L} \nu_{\alpha}\right)\left(\bar{f} \gamma^{\mu} \gamma_{5} f\right)$ & $C_{2,f}^{(6)}$ \\
& $\mathcal{O}_{3}^{(6)}=\left(\overline{\nu^c}_{\beta} P_{L} \nu_{\alpha}\right)\left(\overline{\nu^c}_{\beta'} P_{L} \nu_{\alpha'}\right)^\clubsuit$ & $C_{3}^{(6)}$ \\
& $\mathcal{O}_{4}^{(6)}=\left(\bar{\nu}_{\beta} \gamma_{\mu} P_{L} \nu_{\alpha}\right)\left(\bar{\nu}_{\beta'} \gamma_{\mu} P_{L} \nu_{\alpha'}\right)^\clubsuit$ & $C_{4}^{(6)}$\\
& $\mathcal{O}_{5}^{(6)}=\left(\overline{\nu^c}_{\beta} \sigma^{\mu \nu} P_{L} \nu_{\alpha}\right)\left(\overline{\nu^c}_{\beta'} \sigma^{\mu \nu} P_{L} \nu_{\alpha'}\right)^\clubsuit$ & $C_{5}^{(6)}$\\
\hline
\multirow{9}{*}{\rm dimension-7} & $\mathcal{O}_{1}^{(7)}=\frac{\alpha}{12 \pi}\left(\bar{\nu}_{\beta} P_{L} \nu_{\alpha}\right) F^{\mu \nu} F_{\mu \nu}$ & $C_{1}^{(7)}$ \\
& $\mathcal{O}_{2}^{(7)}=\frac{\alpha}{8 \pi}\left(\bar{\nu}_{\beta} P_{L} \nu_{\alpha}\right) F^{\mu \nu} \widetilde{F}_{\mu \nu}$ & $C_{2}^{(7)}$ \\
& $\mathcal{O}_{5,f}^{(7)}=m_{f}\left(\bar{\nu}_{\beta} P_{L} \nu_{\alpha}\right)(\bar{f} f)$ & $C_{5,f}^{(7)}$ \\
& $\mathcal{O}_{6,f}^{(7)}=m_{f}\left(\bar{\nu}_{\beta} P_{L} \nu_{\alpha}\right)\left(\bar{f} i \gamma_{5} f\right)$ & $C_{6,f}^{(7)}$  \\
& $\mathcal{O}_{7,f}^{(7)}=m_{f}\left(\bar{\nu}_{\beta} \sigma^{\mu \nu} P_{L} \nu_{\alpha}\right)\left(\bar{f} \sigma_{\mu \nu} f\right)$  & $C_{7,f}^{(7)}$ \\
& $\mathcal{O}_{8,f}^{(7)}=\left(\bar{\nu}_{\beta} i\stackrel{\leftrightarrow}{\partial}_{\mu} P_{L} \nu_{\alpha}\right)\left(\bar{f} \gamma^{\mu} f\right)$ & $C_{8,f}^{(7)}$  \\
& $\mathcal{O}_{9,f}^{(7)}=\left(\bar{\nu}_{\beta} i\stackrel{\leftrightarrow}{\partial}_{\mu} P_{L} \nu_{\alpha}\right)\left(\bar{f} \gamma^{\mu}\gamma_5 f\right)$ & $C_{9,f}^{(7)}$  \\
& $\mathcal{O}_{10,f}^{(7)}=\partial_{\mu}\left(\bar{\nu}_{\beta} \sigma^{\mu \nu} P_{L} \nu_{\alpha}\right)\left(\bar{f} \gamma_{\nu} f\right)$ & $C_{10,f}^{(7)}$  \\
& $\mathcal{O}_{11,f}^{(7)}=\partial_{\mu}\left(\bar{\nu}_{\beta} \sigma^{\mu \nu} P_{L} \nu_{\alpha}\right)\left(\bar{f} \gamma_{\nu}\gamma_5 f\right)$ & $C_{11,f}^{(7)}$ \\
\hline
\end{tabular}
\caption{Effective operators relevant for $N_{\rm eff}$ up to dimension-7 with $\alpha,\beta,\alpha',\beta'=e,\mu,\tau$, the neutrino flavor indices, and $f=e$. Operators with $\clubsuit$'s are the extra operators we consider in this work and the symbol ``$c$'' along with related operators means charge conjugation. The last column shows our convention for the Wilson coefficients.}\label{SMEFTNeffOperators}
\end{table}

We start from the SMEFT, obtained by integrating out the heavy new degrees of freedom introduced to the SM, where the Lagrangian can be expressed as the SM Lagrangian, plus a tower of higher-dimension operators $\mathcal{O}^{(j)}$:
\eqal{\mathcal{L}=\mathcal{L}_{\rm SM}+\sum\limits_{j\ge5}\frac{C_j}{\Lambda^{j-4}}\mathcal{O}^{(j)},}
where $C_j$'s are the Wilson coefficients and $\Lambda$ is the characteristic scale of new physics. In this setup, the neutrino masses can be naturally generated through the dimension-5 Weinberg operator\,\cite{Weinberg:1979sa}. However, for the rest of this work, we neglect the masses of neutrinos due to their tininess compared with the other scales involved in our calculation.

In the early Universe where the active fields are neutrinos, photons, electron and positrons, the Universe can be described by the LEFT, where the top quark, the $\rm SU(2)$ gauge bosons, and the Higgs boson have also been integrated out within the SMEFT, inducing both CC and NC neutrino NSIs. See, for example, Ref.\,\cite{Wise:2014oea} for the discussion. However, we point out that CC and NC NSIs are not necessarily generated by heavy particles above the weak scale, instead, it can also be generated by some light particles above the $\mathcal{O}(\rm 100\,MeV)$ scale. To illustrate this point, we briefly discuss a toy $Z'$ and a toy pseudo-scalar models here:
\begin{itemize}
\item The toy $\rm U(1)'$ model we consider, without restricting ourselves to any other constraints such as anomaly cancellation, collider and cosmological constraints etc., is the $Z'$ model that can be written as
\eqal{
\mathcal{L}_{Z'} = \mathcal{L}_{\rm SM} - \frac{1}{4}Z'_{\mu\nu}Z'^{\mu\nu} + \frac{1}{2}m_{Z'}^2 Z'_\mu Z'^\mu - g_{Z'} Z'_\mu \left( \bar{L}\gamma^\mu L + \bar{\ell}_R\gamma^\mu\ell_R \right),
} 
where $L$ and $\ell_R$ are the left-handed lepton doublet and the right-handed lepton singlet under $\rm SU(2)_L$ respectively, and $Z'$ is the new vector boson charged under the $\rm U(1)'$ group. For our purpose, $Z'$ needs not to be above the weak scale, and as long as $Z'$ is above $\sim\mathcal{O}(\rm 100\,MeV)$ or equivalently the muon mass, $Z'$ can be integrated out:
\eqal{
\mathcal{L}_{Z'}\supset &\, \frac{1}{2}m_{Z'}^2 \left(\frac{g_{Z'}}{p^2-m_{Z'}^2}\right)^2\left( \bar{L}\gamma_\mu L + \bar{\ell}_R\gamma_\mu\ell_R \right)\left( \bar{L}\gamma^\mu L + \bar{\ell}_R\gamma^\mu\ell_R \right) \nonumber\\
 &\, - \frac{g_{Z'}^2}{p^2-m_{Z'}^2}\left( \bar{L}\gamma^\mu L + \bar{\ell}_R\gamma^\mu\ell_R \right) \left( \bar{L}\gamma^\mu L + \bar{\ell}_R\gamma^\mu\ell_R \right)\nonumber\\
&\, \xrightarrow[]{p^2\ll m_{Z'}^2}\frac{g_{Z'}^2}{2m_{Z'}^2}\left( \bar{L}\gamma^\mu L + \bar{\ell}_R\gamma^\mu\ell_R \right) \left( \bar{L}\gamma^\mu L + \bar{\ell}_R\gamma^\mu\ell_R \right)+\mathcal{O}\left(\frac{1}{m_{Z'}^4}\right),
}
leading to the NC neutrino-electron and electron-electron contact interactions as seen above when $m_{Z'}^2$ is larger than the momentum transfer $p^2$. During neutrino decoupling, since $p^2$ is of $\mathcal{O}(\rm 10\,MeV)^2$, thus as long as $m_{Z'}$ is above $\mathcal{O}(\rm 100\,MeV)$, the EFT after integrating out $Z'$ serves as a good framework for the study of this new physics.
\item The toy pseudo-scalar model we consider, without considering any theoretical and/or experimental constraints, can be expressed as 
\eqal{
\mathcal{L}_{\rm p.s.} = \mathcal{L}_{\rm SM} + \frac{1}{2}\partial_\mu\phi\partial^\mu\phi-\frac{1}{2}m_{\phi}\phi^2 - i g_\phi^{\alpha\beta} \phi\bar{\nu}_\alpha\gamma_5\nu_\beta, \quad \text{ with } \alpha,\beta=e,\mu,\tau,
}
where $\phi$ is the pseudo-scalar with mass $m_\phi$. Similarly, as long as $m_\phi$ is above $\sim\mathcal{O}(\rm 100\,MeV)$, one can integrate out the particle $\phi$, and obtain the contact neutrino self-interacting operators.
\end{itemize}

The CC NSIs have been recently studied in Refs.\,\cite{Terol-Calvo:2019vck,Du:2020dwr}. In Ref.\,\cite{Du:2020dwr}, the authors took the running and the matching effects at different EFT scales into account, and the resulting constraint on the UV scale $\Lambda$ was found to be as large as about 20\,TeV from neutrino oscillation data. The NC operators are the relevant ones for our study in this work, part of which has been previously studied in Refs.\,\cite{Farzan:2018gtr,Billard:2018jnl,AristizabalSierra:2018eqm,Kosmas:2017tsq,Dent:2017mpr,Liao:2017uzy,Dent:2016wcr,Lindner:2016wff}, and a comprehensive study was recently presented in Ref.\,\cite{Altmannshofer:2018xyo}. Note that, since the authors in Ref.\,\cite{Altmannshofer:2018xyo} were interested in constraints on these NSIs from neutrino experiments, they did not consider any dimension-6 neutrino self-interacting operators as neutrinos only feebly interact with our matter world. However, since these operators are closely related to $N_{\rm eff}$ by modifying the neutrino number and the energy densities directly through neutrino self-interactions, we include these operators in this work and study constraints on these neutrino self-interacting operators from precision measurements of $N_{\rm eff}$.\footnote{Neutrino self-interactions was also proposed to alleviate the Hubble tension between measurements from the Planck\,\cite{Aghanim:2018eyx} and the local groups\,\cite{Riess:2019cxk} in Ref.\,\cite{Kreisch:2019yzn}. For the most recent work on Hubble tension from neutrino self-interactions, see Refs.\,\cite{Brinckmann:2020bcn,Choudhury:2020tka,Das:2020xke,Huang:2021dba}.} We summarize all the EFT operators up to dimension-7 in table\,\ref{SMEFTNeffOperators} following the notations in Ref.\,\cite{Altmannshofer:2018xyo}.

One immediate observation from table\,\ref{SMEFTNeffOperators} is that, the dimension-5 operator in the first row, i.e., the neutrino magnetic dipole operator, corresponds to the light-mediator scenario we discussed at the end of section\,\ref{sec:PSIntegral}. When the intermediate photon shows up in the $s$-channel, there is no IR divergence, and the collision term integral can be calculated analytically. However, since the intermediate photon can also appear in the $t$- and/or $u$-channels for the $\nu_\alpha\nu_\beta\to\gamma^*\to\nu_\alpha\nu_\alpha$\footnote{The collision term integrals for the $\nu_\alpha\nu_\alpha\to\gamma^*\to\nu_\alpha\nu_\alpha$ process simply vanish since the initial and the final states have exactly the same temperature, thus the number density and the energy densities of $\nu_\alpha$ remain the same before and after the interaction.} process for example, the collision term integrals would exhibit the IR divergence discussed earlier. In principle, one could remove this divergence by applying the KLN theorem or following the procedure discussed in Ref.\,\cite{Frye:2018xjj} for any inclusive observables. However, since the scenario with a light mediator resides in a different regime compared with all the other operators listed in table\,\ref{SMEFTNeffOperators}, we leave the light mediator scenario for a future project. We also point out that this operator is very stringently constrained from the magnetic moment of $\nu_e$ using Borexino Phase-II solar neutrino data\,\cite{Altmannshofer:2018xyo,Borexino:2017fbd}, which justifies our ignorance of the $\mathcal{O}_1^{(5)}$ operator for the calculation of $N_{\rm eff}$. We will discuss more on this in section\,\ref{sec:EFTConstraints}.

Now, including corrections from the dimension-6 and dimension-7 operators in table\,\ref{SMEFTNeffOperators}, the invariant amplitude $\langle\mathcal{M}^{2}\rangle_{1+2\to3+4}$ in eq.\,\eqref{CollIntFinal} can generically be written as:
\eqal{\langle\mathcal{M}^{2}\rangle_{1+2\to3+4} = \langle \mathcal{M}_{\rm SM}^{2} + \mathcal{M}_{\rm EFT}^{2} + 2\,{\rm Re} \mathcal{M}_{\rm SM}\cdot\mathcal{M}_{\rm EFT}^\dagger\rangle_{1+2\to3+4},\label{amp2Inter}}
where $\mathcal{M}_{\rm SM}$ and $\mathcal{M}_{\rm EFT}$ are the amplitudes from $\mathcal{L}_{\rm SM}$ and the EFT operators in table\,\ref{SMEFTNeffOperators} respectively. Clearly, when the potential new physics scale $\Lambda$ is about or above $\mathcal{O}(\Lambda_W)$ with $\Lambda_W$ the weak scale, the interference term in eq.\,\eqref{amp2Inter} would be of the same order as the SM contributions, and the $\langle\mathcal{M}_{\rm EFT}^{2}\rangle$ term can then be safely neglected.

However, we point out that in the case where $\Lambda\ll\Lambda_W$, though contributions from the EFT operators dominate, one can not simply discard the $\langle\mathcal{M}_{\rm SM}^{2}\rangle$ term in eq.\,\eqref{amp2Inter} since for some of the operators in table\,\ref{SMEFTNeffOperators}, for example, the $\mathcal{O}_{3,4,5}^{6}$ operators, the $\langle\mathcal{M}_{\rm SM}^{2}\rangle$ is the only part that tells how $T_{\gamma}$ and $T_{\nu_\alpha}$ evolve with time as we will see below. In light of this, we always keep all the three terms in eq.\,\eqref{amp2Inter} during our calculation, while using the large $\Lambda$ limit to cross check our results.

By plugging in $\mathcal{M}_{\rm SM}$ and $\mathcal{M}_{\rm EFT}$ in eq.\,\eqref{amp2Inter}, one obtains $\langle\mathcal{M}^{2}\rangle_{1+2\to3+4}$ as a function of $p_{ij}$, the SM model parameters, the scale of new physics $\Lambda$, and the Wilson coefficients $C_{j}$ in the last column of table\,\ref{SMEFTNeffOperators}. One can then calculate the collision term integrals through finishing the integral shown in eq.\,\eqref{CollIntFinal}. However, due to momentum-energy conservation, redundancy exists in the calculation of collision term integrals. This redundancy can be sufficiently removed by first choosing a set of basis, which we discuss in the next subsection.

\subsection{Choices of the independent bases}\label{sec:EFTOpeBas}
To start, we realize that for both the SM and the EFT contributions, the relevant processes are either $2\to2$ scattering or $2\to2$  annihilating processes. Denoting these processes generically as $1+2\to3+4$ with momentum $p_i$ for the $i$-th particle, the invariant amplitude $\langle\mathcal{M}^2\rangle_{1+2\to3+4}$, defined in eq.\,\eqref{amp2Inter}, can be expressed as a tower of the momentum scalar product $p_{ij}$ defined in eq.\,\eqref{def:pij}:
\eqal{
\langle\mathcal{M}^2\rangle_{1+2\to3+4}=\sum\limits_{i,\cdots,t=1}^{4}\sum\limits_{k=0}^{\infty}c_{ij\cdots mn\cdots st}(\{m\},\{g\}, \{C\}, \Lambda, \{T\}, \{\mu\})\cdot \underbrace{p_{ij}\cdots p_{mn}\cdots p_{st}}_{\text{k = number of }p_{ij}\text{'s}},\label{eq:ampgeneral}
}
where the coefficients $c_{ij\cdots mn\cdots st}$'s are generically functions of the mass set $\{m\}$ and the coupling set $\{g\}$ of the SM, the new physics scale $\Lambda$ and the corresponding Wilson coefficient set $\{C\}$ in the last column of table\,\ref{SMEFTNeffOperators}, the temperatures $T_{\gamma,\nu_\alpha}$ and the chemical potentials $\mu_{\gamma,\nu_\alpha}$.

\begin{table}[t]
 \renewcommand{\arraystretch}{1.3}
\centering
\begin{tabular}{|c|c|c|}
\hline
$k$ & Bases & Number of bases \\
\hline
0 & 1 & 1\\
\hline
1 & $p_{12},\quad p_{13},\quad p_{14}$ & 3 \\
\hline
\multirow{1}{*}{2} & $ p_{12}^2,\quad  p_{12}\cdot p_{13},\quad  p_{12}\cdot p_{14},\quad p_{13}^2,\quad  p_{13}\cdot p_{14},\quad  p_{14}^2   $ & 6\\
\hline
\multirow{2}{*}{3} & $p_{12}^3,\quad p_{12}^2 \cdot p_{13},\quad p_{12}^2\cdot p_{14},\quad p_{12}\cdot p_{13}^2,\quad p_{12}\cdot p_{13}\cdot p_{14},\quad p_{12}\cdot p_{14}^2$ & \multirow{2}{*}{10} \\
 & $p_{13}^3,\quad p_{13}^2\cdot p_{14},\quad p_{13}\cdot p_{14}^2,\quad p_{14}^3 $ & \\
\hline
\end{tabular}
\caption{The bases we choose with different $k$'s for the calculation of collision term integrals. See the main text for more discussion.}\label{AMP2IndepBases}
\end{table}

In this work, since we are only interested in contributions from the SM and the EFT operators up to dimension-7 as listed in table\,\ref{SMEFTNeffOperators}, it turns out that we only need to consider $k$'s up to $k=3$ in eq.\,\eqref{eq:ampgeneral}, which then gives, by leaving out the arguments of $c_{ij\cdots mn\cdots st}$'s, 
\eqal{
\langle\mathcal{M}^2\rangle_{1+2\to3+4}=&\,c_{0}\nonumber\\
&+\sum\limits_{\substack{i,j=1 \\ i\ne j}}^4 c_{ij}\cdot p_{ij} + \sum\limits_{\substack{i,\dots,n=1 \\ i\ne j\\m\ne n}}^4 c_{ijmn}\cdot p_{ij}\cdot p_{mn}\nonumber\\
&+\sum\limits_{\substack{i,\cdots,t=1 \\ i\ne j\\ m\ne n\\ s\ne t}}^4 c_{ijmnst}
\cdot {p_{ij}\cdot p_{mn}\cdot p_{st}}.\label{AMP2Decom}
}
Note that for $k=1$, and similarly for the $k=2, 3$ cases, we have included contributions from $i=j$ in the $c_0$ term from the on-shell conditions. On the other hand, terms in the second and the third lines of eq.\,\eqref{AMP2Decom} are not all independent from momentum-energy conservation, resulting in redundancy when one calculates the collision term integrals from eqs.\,\eqref{AMP2Decom} and \eqref{CollIntFinal}. However, this redundancy can be sufficiently removed by first choosing an independent basis in terms of $p_{ij}$, and then rewrite $\langle\mathcal{M}^2\rangle_{1+2\to3+4}$ as a linear combination of these bases. Depending on $k$, the independent bases we choose are presented in table\,\ref{AMP2IndepBases}. One can then readily express the momentum tower as linear combinations of these bases. For example,
\eqal{p_{12}\cdot p_{23}\cdot p_{24} = &\,\frac{1}{4}\left( \left(m_1^2 - m_2^2\right)^2 - \left( m_3^2 - m_4^2 \right)^2 \right)p_{12} + \frac{1}{2} \left( m_2^2 + m_3^2 - m_1^2 - m_4^2 \right)p_{12}\cdot p_{13} \nonumber\\
& + \frac{1}{2} \left( m_2^2 + m_4^2 - m_1^2 - m_3^2 \right)p_{12}\cdot p_{14} + p_{12}\cdot p_{13}\cdot p_{14}\\
\to&\, p_{12}\cdot p_{13}\cdot p_{14}\quad\quad{\text{in the massless limit.}}}

At this stage, the collision term integral in eq.\,\eqref{CollIntFinal} boils down to the collision term integral with $\langle\mathcal{M}^2\rangle_{1+2\to3+4}$ being replaced by the independent bases shown in table\,\ref{AMP2IndepBases}. Particularly, when all the masses involved are vanishing, $\langle\mathcal{M}^2\rangle_{1+2\to3+4}$ generically simplifies significantly as seen from the example above. Therefore, as also discussed at the end of section\,\ref{sec:PSIntegral}, to obtain analytical results for the collision term integrals, we assume $m_i=0\,(i=1,\dots,4)$\footnote{Since the electron mass $m_e$ is the only one matters here, this assumption basically means that, when calculating the collision term integrals in eq.\,\eqref{CollIntFinal}, we take the untra-relativistic limit for electrons in the early Universe.} and all particles obey the Maxwell-Boltzmann distribution. We then present the complete generic and analytical dictionary of the collision term integrals in section\,\ref{sec:ComDictColl}. Corrections from finite electron mass $m_e$, {\color{black}spin statistics and neutrino chemical potentials} are discussed in section\,\ref{sec:meandSpinCorrections}.

\subsection{Corrections from $m_e$, spin statistics and chemical potentials}\label{sec:meandSpinCorrections}
As already noticed in Ref.\,\cite{Escudero:2020dfa}, corrections from finite electron mass $m_e$ and spin-statistics have to be included to reproduce $N_{\rm eff}$ obtained from the density matrix formalism. Furthermore, as one can see from Table\,1 of Ref.\,\cite{Escudero:2020dfa}, these corrections are of the same order as the finite temperature corrections discussed at the beginning of this section. Thus, to be consistent, these corrections have to be included. 

In Ref.\,\cite{Escudero:2020dfa}, finite $m_e$ corrections are obtained by finding the ratios of the collision term integrals in eq.\,\eqref{CollIntFinal} by switching on and off $m_e$. Similarly, corrections from spin statistics are computed by finding the ratios of the collision term integrals with Fermi-Dirac/Bose-Einstein and Maxwell-Boltzmann distributions respectively. For a detailed discussion, see Refs.\,\cite{Escudero:2020dfa,Luo:2020sho}. Though the methods used for the collision term integrals are different, we reproduce the numbers in Table 6 of Ref.\,\cite{Escudero:2020dfa} and/or Table III of Ref.\,\cite{Luo:2020sho}. These corrections are then included in the Boltzmann equations and used to solve $N_{\rm eff}$. The results are presented and discussed in more detail in section\,\ref{sec:EFTConstraints}.

{\color{black}We also comment on that neutrino chemical potentials are highly suppressed due to the rapid $\bar{\nu}\nu\leftrightarrow e^+e^-\leftrightarrow\gamma\gamma$ conversion and that the electron chemical potential is negligibly small compared to the plasma temperature in the early Universe. In our setup, in order to be generic, we keep neutrino chemical potentials throughout our analytical calculations, but stress that they would have no visible impact on the current/planned precision measurement of $N_{\rm eff}$. For our numerical calculations, we choose $\mu_\nu=\mu_{\bar{\nu}}$ and $|\mu_\nu/T_\gamma|=10^{-4}$ with $T_\gamma=T_\nu=10\rm\,MeV$ as our initial conditions, and verify that this setup is numerically equivalent to vanishing neutrino chemical potentials as expected.}

\subsection{A complete generic and analytical dictionary of the collision term integrals}\label{sec:ComDictColl}

In last subsection, we list in table\,\ref{AMP2IndepBases} the independent bases by which the invariant amplitudes $\langle\mathcal{M}^2\rangle_{1+2\to3+4}$ can be expressed, and conclude that the redundancy of collision term integrals from momentum-energy conservation can be removed by working with these bases directly. In this subsection, we provide the complete analytical dictionary of the collision term integrals for particle ``1'' and up to $k=3$, with $k$ the number of $p_{ij}$'s in the invariant amplitude. We note that a subset of this complete dictionary was presented in the appendices of Ref.\,\cite{Escudero:2018mvt,Luo:2020sho}, which agrees with our results presented in this subsection as long as one specifies $T_i$ and $\mu_i$ accordingly.

For the collision term integrals, we follow the procedure briefly summarized in section\,\ref{sec:PSIntegral} and stick to our notation in eq.\,\eqref{CollIntFinal}, where $j=0$ represents the collision term integral for the number density and $j=1$ that for the energy density. The dependence on $p_{ij}$ of $\langle\mathcal{M}^{2}\rangle_{1+2\to3+4}$ is reflected by the argument of $C^{(j)}$.\footnote{We stress that the argument of $C^{(j)}$ here is only used to reflect the dependence on $p_{ij}$ of $\langle\mathcal{M}^{2}\rangle_{1+2\to3+4}$, and this argument does not mean that $C^{(j)}$ depends on $p_{ij}$. Instead, $C^{(j)}$ only depends on the model parameters, the new physics scale, the Wilson coefficients, the temperatures $T_{\gamma,\nu_\alpha}$ and the chemical potentials $\mu_{\nu_\alpha}$.} For example, $C^{(0)}(p_{12})$ means the collision term integral for the number density with $\langle\mathcal{M}^{2}\rangle_{1+2\to3+4}=p_{12}$\footnote{We leave out any overall factors in $\langle\mathcal{M}^{2}\rangle_{1+2\to3+4}$ that are independent of $p_{ij}$ here and in the following.} in eq.\,\eqref{CollIntFinal}.

\subsubsection{$\langle\mathcal{M}^{2}\rangle_{1+2\to3+4}=1$}
\eqal{
C^{(j)} (1)= \left\{\begin{array}{ll} \frac{1}{128 \pi^{5}}\left[-{e}^{\frac{\mu_1}{T_1}+\frac{\mu_2}{T_2}} T_1^{2}T_2^{2} + {e}^{\frac{\mu_3}{T_3}+\frac{\mu_4}{T_4}} T_3^{2} T_4^{2}\right], & j=0 \\ & \\ \frac{1}{128 \pi^{5}}\left[-2{e}^{\frac{\mu_1}{T_1}+\frac{\mu_2}{T_2}} T_1^{3}T_2^{2} + {e}^{\frac{\mu_3}{T_3}+\frac{\mu_4}{T_4}} T_3^{2} T_4^{2}(T_3+T_4)\right], & j=1 \end{array}\,\,,\right.\label{coll:ampconst}
}
From eq.\,\eqref{coll:ampconst}, one notes that in the $j=0$ case, the collision term integral, corresponding to the number density, vanishes when $T_1=T_3$ and $T_2=T_4$. This is expected since it actually stands for a scattering process where the number density for each species remains the same before and after the interaction. This conclusion holds generically and is independent of the form of $\langle\mathcal{M}^{2}\rangle_{1+2\to3+4}$, as one can also see clearly from the results below. On the other hand, if $T_1=T_2=T_3=T_4$ and $\mu_1=\mu_2=\mu_3=\mu_4$, then all the $C^{(j)}$'s vanish, which is also as expected since particle self interactions do not modify the number and the energy densities as long as thermal equilibrium is maintained. This observation also acts as a cross-check of our analytical results presented in these subsections.

\subsubsection{$\langle\mathcal{M}^{2}\rangle_{1+2\to3+4}=p_{ij}$}
\eqal{
C^{(j)}(p_{12}) = &\, \left\{\begin{array}{ll} \frac{1}{32 \pi^{5}}\left[-{e}^{\frac{\mu_1}{T_1}+\frac{\mu_2}{T_2}} T_1^{3}T_2^{3} + {e}^{\frac{\mu_3}{T_3}+\frac{\mu_4}{T_4}} T_3^{3} T_4^{3}\right], & j=0 \\ & \\ \frac{1}{64 \pi^{5}}\left[-6{e}^{\frac{\mu_1}{T_1}+\frac{\mu_2}{T_2}} T_1^{4}T_2^{3} + 3{e}^{\frac{\mu_3}{T_3}+\frac{\mu_4}{T_4}} T_3^{3} T_4^{3}(T_3+T_4)\right], & j=1 \end{array}\,\,,\right.\\
C^{(j)}(p_{13}) =&\,  \left\{\begin{array}{ll} \frac{1}{64 \pi^{5}}\left[-{e}^{\frac{\mu_1}{T_1}+\frac{\mu_2}{T_2}} T_1^{3}T_2^{3} + {e}^{\frac{\mu_3}{T_3}+\frac{\mu_4}{T_4}} T_3^{3} T_4^{3}\right], & j=0 \\ & \\ \frac{1}{64 \pi^{5}}\left[-3{e}^{\frac{\mu_1}{T_1}+\frac{\mu_2}{T_2}} T_1^{4}T_2^{3} + {e}^{\frac{\mu_3}{T_3}+\frac{\mu_4}{T_4}} T_3^{3} T_4^{3}(T_3+2T_4)\right], & j=1 \end{array}\,\,,\right.\\
C^{(j)}(p_{14}) =&\,  \left\{\begin{array}{ll} \frac{1}{64 \pi^{5}}\left[-{e}^{\frac{\mu_1}{T_1}+\frac{\mu_2}{T_2}} T_1^{3}T_2^{3} + {e}^{\frac{\mu_3}{T_3}+\frac{\mu_4}{T_4}} T_3^{3} T_4^{3}\right], & j=0 \\ & \\ \frac{1}{64 \pi^{5}}\left[-3{e}^{\frac{\mu_1}{T_1}+\frac{\mu_2}{T_2}} T_1^{4}T_2^{3} + {e}^{\frac{\mu_3}{T_3}+\frac{\mu_4}{T_4}} T_3^{3} T_4^{3}(2T_3+T_4)\right], & j=1 \end{array}\,\,,\right.
}

\subsubsection{$\langle\mathcal{M}^{2}\rangle_{1+2\to3+4}=p_{ij}\cdot p_{mn}$}
\eqal{
C^{(j)}(p_{12}^2) = &\, \left\{\begin{array}{ll} \frac{3}{8 \pi^{5}}\left[-{e}^{\frac{\mu_1}{T_1}+\frac{\mu_2}{T_2}} T_1^{4}T_2^{4} + {e}^{\frac{\mu_3}{T_3}+\frac{\mu_4}{T_4}} T_3^{4} T_4^{4}\right], & j=0 \\ & \\ \frac{1}{4 \pi^{5}}\left[-6{e}^{\frac{\mu_1}{T_1}+\frac{\mu_2}{T_2}} T_1^{5}T_2^{4} + 3{e}^{\frac{\mu_3}{T_3}+\frac{\mu_4}{T_4}} T_3^{4} T_4^{4}(T_3+T_4)\right], & j=1 \end{array}\,\,,\right.\label{Cjresp12SQ}\\
C^{(j)}(p_{12}\cdot p_{13}) = &\, \left\{\begin{array}{ll} \frac{3}{16 \pi^{5}}\left[-{e}^{\frac{\mu_1}{T_1}+\frac{\mu_2}{T_2}} T_1^{4}T_2^{4} + {e}^{\frac{\mu_3}{T_3}+\frac{\mu_4}{T_4}} T_3^{4} T_4^{4}\right], & j=0 \\ & \\ \frac{1}{4 \pi^{5}}\left[-3{e}^{\frac{\mu_1}{T_1}+\frac{\mu_2}{T_2}} T_1^{5}T_2^{4} + {e}^{\frac{\mu_3}{T_3}+\frac{\mu_4}{T_4}} T_3^{4} T_4^{4}(T_3+2T_4)\right], & j=1 \end{array}\,\,,\right.\\
C^{(j)}(p_{12}\cdot p_{14}) = &\, \left\{\begin{array}{ll} \frac{3}{16 \pi^{5}}\left[-{e}^{\frac{\mu_1}{T_1}+\frac{\mu_2}{T_2}} T_1^{4}T_2^{4} + {e}^{\frac{\mu_3}{T_3}+\frac{\mu_4}{T_4}} T_3^{4} T_4^{4}\right], & j=0 \\ & \\ \frac{1}{4 \pi^{5}}\left[-3{e}^{\frac{\mu_1}{T_1}+\frac{\mu_2}{T_2}} T_1^{5}T_2^{4} + {e}^{\frac{\mu_3}{T_3}+\frac{\mu_4}{T_4}} T_3^{4} T_4^{4}(2T_3+T_4)\right], & j=1 \end{array}\,\,,\right.\\
C^{(j)}(p_{13}^2) = &\, \left\{\begin{array}{ll} \frac{1}{8 \pi^{5}}\left[-{e}^{\frac{\mu_1}{T_1}+\frac{\mu_2}{T_2}} T_1^{4}T_2^{4} + {e}^{\frac{\mu_3}{T_3}+\frac{\mu_4}{T_4}} T_3^{4} T_4^{4}\right], & j=0 \\ & \\ \frac{1}{8 \pi^{5}}\left[-4{e}^{\frac{\mu_1}{T_1}+\frac{\mu_2}{T_2}} T_1^{5}T_2^{4} + {e}^{\frac{\mu_3}{T_3}+\frac{\mu_4}{T_4}} T_3^{4} T_4^{4}(T_3+3T_4)\right], & j=1 \end{array}\,\,,\right.\label{Cjresp13SQ}\\
C^{(j)}(p_{13}\cdot p_{14}) = &\, \left\{\begin{array}{ll} \frac{1}{16 \pi^{5}}\left[-{e}^{\frac{\mu_1}{T_1}+\frac{\mu_2}{T_2}} T_1^{4}T_2^{4} + {e}^{\frac{\mu_3}{T_3}+\frac{\mu_4}{T_4}} T_3^{4} T_4^{4}\right], & j=0 \\ & \\ \frac{1}{8 \pi^{5}}\left[-2{e}^{\frac{\mu_1}{T_1}+\frac{\mu_2}{T_2}} T_1^{5}T_2^{4} + {e}^{\frac{\mu_3}{T_3}+\frac{\mu_4}{T_4}} T_3^{4} T_4^{4}(T_3+T_4)\right], & j=1 \end{array}\,\,,\right.\\
C^{(j)}(p_{14}^2) = &\, \left\{\begin{array}{ll} \frac{1}{8 \pi^{5}}\left[-{e}^{\frac{\mu_1}{T_1}+\frac{\mu_2}{T_2}} T_1^{4}T_2^{4} + {e}^{\frac{\mu_3}{T_3}+\frac{\mu_4}{T_4}} T_3^{4} T_4^{4}\right], & j=0 \\ & \\ \frac{1}{8 \pi^{5}}\left[-4{e}^{\frac{\mu_1}{T_1}+\frac{\mu_2}{T_2}} T_1^{5}T_2^{4} + {e}^{\frac{\mu_3}{T_3}+\frac{\mu_4}{T_4}} T_3^{4} T_4^{4}(3T_3+T_4)\right], & j=1 \end{array}\,\,,\right.\label{Cjresp14SQ}
}

\subsubsection{$\langle\mathcal{M}^{2}\rangle_{1+2\to3+4}=p_{ij}\cdot p_{mn}\cdot p_{st}$}
\eqal{
C^{(j)}(p_{12}^3) = &\, \left\{\begin{array}{ll} -\frac{9}{\pi^{5}}\left[{e}^{\frac{\mu_1}{T_1}+\frac{\mu_2}{T_2}} T_1^{5}T_2^{5} - {e}^{\frac{\mu_3}{T_3}+\frac{\mu_4}{T_4}} T_3^{5} T_4^{5}\right], & j=0 \\ & \\ -\frac{45}{2 \pi^{5}}\left[2{e}^{\frac{\mu_1}{T_1}+\frac{\mu_2}{T_2}} T_1^{6}T_2^{5} - {e}^{\frac{\mu_3}{T_3}+\frac{\mu_4}{T_4}} T_3^{5} T_4^{5}(T_3+T_4)\right], & j=1 \end{array}\,\,,\right.\\
C^{(j)}(p_{12}^2\cdot p_{13}) = &\, \left\{\begin{array}{ll} -\frac{9}{2\pi^{5}}\left[{e}^{\frac{\mu_1}{T_1}+\frac{\mu_2}{T_2}} T_1^{5}T_2^{5} - {e}^{\frac{\mu_3}{T_3}+\frac{\mu_4}{T_4}} T_3^{5} T_4^{5}\right], & j=0 \\ & \\ -\frac{15}{2 \pi^{5}}\left[3{e}^{\frac{\mu_1}{T_1}+\frac{\mu_2}{T_2}} T_1^{6}T_2^{5} - {e}^{\frac{\mu_3}{T_3}+\frac{\mu_4}{T_4}} T_3^{5} T_4^{5}(T_3+2T_4)\right], & j=1 \end{array}\,\,,\right.\\
C^{(j)}(p_{12}^2\cdot p_{14}) = &\, \left\{\begin{array}{ll} -\frac{9}{2\pi^{5}}\left[{e}^{\frac{\mu_1}{T_1}+\frac{\mu_2}{T_2}} T_1^{5}T_2^{5} - {e}^{\frac{\mu_3}{T_3}+\frac{\mu_4}{T_4}} T_3^{5} T_4^{5}\right], & j=0 \\ & \\ -\frac{15}{2 \pi^{5}}\left[3{e}^{\frac{\mu_1}{T_1}+\frac{\mu_2}{T_2}} T_1^{6}T_2^{5} - {e}^{\frac{\mu_3}{T_3}+\frac{\mu_4}{T_4}} T_3^{5} T_4^{5}(2T_3+T_4)\right], & j=1 \end{array}\,\,,\right.
}
\eqal{
C^{(j)}(p_{12}\cdot p_{13}^2) = &\, \left\{\begin{array}{ll} -\frac{3}{\pi^{5}}\left[{e}^{\frac{\mu_1}{T_1}+\frac{\mu_2}{T_2}} T_1^{5}T_2^{5} - {e}^{\frac{\mu_3}{T_3}+\frac{\mu_4}{T_4}} T_3^{5} T_4^{5}\right], & j=0 \\ & \\ -\frac{15}{4 \pi^{5}}\left[4{e}^{\frac{\mu_1}{T_1}+\frac{\mu_2}{T_2}} T_1^{6}T_2^{5} - {e}^{\frac{\mu_3}{T_3}+\frac{\mu_4}{T_4}} T_3^{5} T_4^{5}(T_3+3T_4)\right], & j=1 \end{array}\,\,,\right.\\
C^{(j)}(p_{12}\cdot p_{13}\cdot p_{14}) = &\, \left\{\begin{array}{ll} \frac{3}{2\pi^{5}}\left[-{e}^{\frac{\mu_1}{T_1}+\frac{\mu_2}{T_2}} T_1^{5}T_2^{5} + {e}^{\frac{\mu_3}{T_3}+\frac{\mu_4}{T_4}} T_3^{5} T_4^{5}\right], & j=0 \\ & \\ \frac{15}{4 \pi^{5}}\left[-2{e}^{\frac{\mu_1}{T_1}+\frac{\mu_2}{T_2}} T_1^{6}T_2^{5} + {e}^{\frac{\mu_3}{T_3}+\frac{\mu_4}{T_4}} T_3^{5} T_4^{5}(T_3+T_4)\right], & j=1 \end{array}\,\,,\right.\\
C^{(j)}(p_{12}\cdot p_{14}^2) = &\, \left\{\begin{array}{ll} -\frac{3}{\pi^{5}}\left[{e}^{\frac{\mu_1}{T_1}+\frac{\mu_2}{T_2}} T_1^{5}T_2^{5} - {e}^{\frac{\mu_3}{T_3}+\frac{\mu_4}{T_4}} T_3^{5} T_4^{5}\right], & j=0 \\ & \\ -\frac{15}{4 \pi^{5}}\left[4{e}^{\frac{\mu_1}{T_1}+\frac{\mu_2}{T_2}} T_1^{6}T_2^{5} - {e}^{\frac{\mu_3}{T_3}+\frac{\mu_4}{T_4}} T_3^{5} T_4^{5}(3T_3+T_4)\right], & j=1 \end{array}\,\,,\right.\\
C^{(j)}(p_{13}^3) = &\, \left\{\begin{array}{ll} \frac{9}{4\pi^{5}}\left[-{e}^{\frac{\mu_1}{T_1}+\frac{\mu_2}{T_2}} T_1^{5}T_2^{5} + {e}^{\frac{\mu_3}{T_3}+\frac{\mu_4}{T_4}} T_3^{5} T_4^{5}\right], & j=0 \\ & \\ \frac{9}{4 \pi^{5}}\left[-5{e}^{\frac{\mu_1}{T_1}+\frac{\mu_2}{T_2}} T_1^{6}T_2^{5} + {e}^{\frac{\mu_3}{T_3}+\frac{\mu_4}{T_4}} T_3^{5} T_4^{5}(T_3+4T_4)\right], & j=1 \end{array}\,\,,\right.\\
C^{(j)}(p_{13}^2\cdot p_{14}) = &\, \left\{\begin{array}{ll} -\frac{3}{4\pi^{5}}\left[{e}^{\frac{\mu_1}{T_1}+\frac{\mu_2}{T_2}} T_1^{5}T_2^{5} - {e}^{\frac{\mu_3}{T_3}+\frac{\mu_4}{T_4}} T_3^{5} T_4^{5}\right], & j=0 \\ & \\ -\frac{3}{4 \pi^{5}}\left[5{e}^{\frac{\mu_1}{T_1}+\frac{\mu_2}{T_2}} T_1^{6}T_2^{5} - {e}^{\frac{\mu_3}{T_3}+\frac{\mu_4}{T_4}} T_3^{5} T_4^{5}(2T_3+3T_4)\right], & j=1 \end{array}\,\,,\right.\\
C^{(j)}(p_{13}\cdot p_{14}^2) = &\, \left\{\begin{array}{ll} -\frac{3}{4\pi^{5}}\left[{e}^{\frac{\mu_1}{T_1}+\frac{\mu_2}{T_2}} T_1^{5}T_2^{5} - {e}^{\frac{\mu_3}{T_3}+\frac{\mu_4}{T_4}} T_3^{5} T_4^{5}\right], & j=0 \\ & \\ -\frac{3}{4 \pi^{5}}\left[5{e}^{\frac{\mu_1}{T_1}+\frac{\mu_2}{T_2}} T_1^{6}T_2^{5} - {e}^{\frac{\mu_3}{T_3}+\frac{\mu_4}{T_4}} T_3^{5} T_4^{5}(3T_3+2T_4)\right], & j=1 \end{array}\,\,,\right.\\
C^{(j)}(p_{14}^3) = &\, \left\{\begin{array}{ll} -\frac{9}{4\pi^{5}}\left[{e}^{\frac{\mu_1}{T_1}+\frac{\mu_2}{T_2}} T_1^{5}T_2^{5} - {e}^{\frac{\mu_3}{T_3}+\frac{\mu_4}{T_4}} T_3^{5} T_4^{5}\right], & j=0 \\ & \\ -\frac{9}{4 \pi^{5}}\left[5{e}^{\frac{\mu_1}{T_1}+\frac{\mu_2}{T_2}} T_1^{6}T_2^{5} - {e}^{\frac{\mu_3}{T_3}+\frac{\mu_4}{T_4}} T_3^{5} T_4^{5}(4T_3+T_4)\right], & j=1 \end{array}\,\,,\right.
}

With this complete dictionary, one can readily write down the Boltzmann equations for $T_{\gamma}$ and $T_\nu$ as long as $\langle\mathcal{M}^{2}\rangle_{1+2\to3+4}$ in eqs.\,(\ref{TvFinal}-\ref{TgammaFinal}) is known. For example, besides SM contributions, only the $\mathcal{O}_{3,4,5}^{(6)}$ operators listed in table\,\ref{SMEFTNeffOperators} introduce new neutrino self-interactions and thus modify the number and the energy densities of neutrinos of different flavors. For the $\nu_\alpha\nu_\beta\to\nu_\alpha\nu_\beta$ ($\alpha\ne\beta$) process, we find
\small{
\eqal{
\langle\mathcal{M}^{2}\rangle_{\nu_\alpha\nu_\beta\to\nu_\alpha\nu_\beta}^{{\rm SM}+\mathcal{O}_{3,4,5}^{(6)}} = &\, \left[ 32G_F^2\cdot p_{12}^2+ \frac{32\sqrt{2}G_FC_{4}^{(6)}}{\Lambda^2} \cdot p_{12}^2 + \frac{16}{\Lambda^4}\left((C_{4}^{(6)})^2 -2(C_{3}^{(6)}-4C_{5}^{(6)})C_{5}^{(6)}\right) \cdot p_{12}^2 \right.\nonumber\\
&\left. + \frac{4}{\Lambda^4} \left((C_{3}^{(6)})^2 -16(C_{5}^{(6)})^2\right) \cdot p_{13}^2 + \frac{32C_{5}^{(6)}}{\Lambda^4} \left( C_{3}^{(6)} + 4 C_{5}^{(6)} \right) \cdot p_{14}^2 \right],
}}where $G_F$ is the Fermi constant and $\Lambda$ is the scale of the potential new physics. The first term in the square bracket is the pure SM contributions, the second term is the interference term between the SM and the $\mathcal{O}_{4}^{(6)}$ operator, and the remaining terms are the pure contributions from the $\mathcal{O}_{3,4,5}^{(6)}$ operators. One can then immediately write down the collision term integrals for the $\nu_\alpha\nu_\beta\to\nu_\alpha\nu_\beta$ process as
\small{
\eqal{
C^{(j)}_{\nu_\alpha\nu_\beta\to\nu_\alpha\nu_\beta} = &\, \left[ 32G_F^2\cdot p_{12}^2+ \frac{32\sqrt{2}G_FC_{4}^{(6)}}{\Lambda^2} \cdot C^{(j)}(p_{12}^2) \right.\nonumber\\
&\quad + \frac{16}{\Lambda^4}\left((C_{4}^{(6)})^2 -2(C_{3}^{(6)}-4C_{5}^{(6)})C_{5}^{(6)}\right) \cdot C^{(j)}(p_{12}^2)\nonumber\\
&\quad + \frac{4}{\Lambda^4} \left((C_{3}^{(6)})^2 -16(C_{5}^{(6)})^2\right) \cdot C^{(j)}(p_{13}^2) \nonumber\\
&\left. \quad + \frac{32C_{5}^{(6)}}{\Lambda^4} \left( C_{3}^{(6)} + 4 C_{5}^{(6)} \right) \cdot C^{(j)}(p_{14}^2) \right],\label{eq:colexample}
}
}where $C^{(j)}(p_{12}^2)$, $C^{(j)}(p_{13}^2)$ and $C^{(j)}(p_{14}^2)$ are given in eqs.\,\eqref{Cjresp12SQ}, \eqref{Cjresp13SQ} and \eqref{Cjresp14SQ} respectively, and $j=0\,(1)$ is for the number (energy) density of $\nu_{\alpha}$. Though not shown explicitly, $C^{(j)}(p_{12}^2)$, $C^{(j)}(p_{13}^2)$ and $C^{(j)}(p_{14}^2)$ depend on the temperatures and the chemical potentials of $\nu_{\alpha,\beta}$. Specifically, one has $T_{1,3}=T_{\nu_\alpha}$, $T_{2,4}=T_{\nu_\beta}$, $\mu_{1,3}=\mu_{\nu_\alpha}$, $\mu_{2,4}=\mu_{\nu_\beta}$ for $\alpha,\beta=e,\mu,\tau$ and $\alpha\ne\beta$.

The complete results of $\langle\mathcal{M}^{2}\rangle$ from SM and the EFT operators listed in table\,\ref{SMEFTNeffOperators} are given in an auxiliary {\tt Mathematica} notebook file for all relevant processes, together with all the replacement rules to rewrite $\langle\mathcal{M}^{2}\rangle$ in terms of the bases listed in table\,\ref{AMP2IndepBases}.\footnote{The replacement rules are obtained with the help of {\tt Package-X}\,\cite{Patel:2016fam}.} We point out that when all the Wilson coefficients vanish therein, we reproduce the SM results as presented in, for example, Ref.\,\cite{Dolgov:2002wy}. Using the complete dictionary summarized in this section and building our code upon {\tt nudec\_BSM} from Ref.\,\cite{Escudero:2020dfa}, we study corrections to $N_{\rm eff}$ from the NSI operators in table\,\ref{SMEFTNeffOperators}, and discuss the results in section\,\ref{sec:EFTConstraints}.

\section{Constraints on EFT operators from $N_{\rm eff}$}\label{sec:EFTConstraints}
  \begin{figure}[t]
        \center{\includegraphics[width=\textwidth]
        {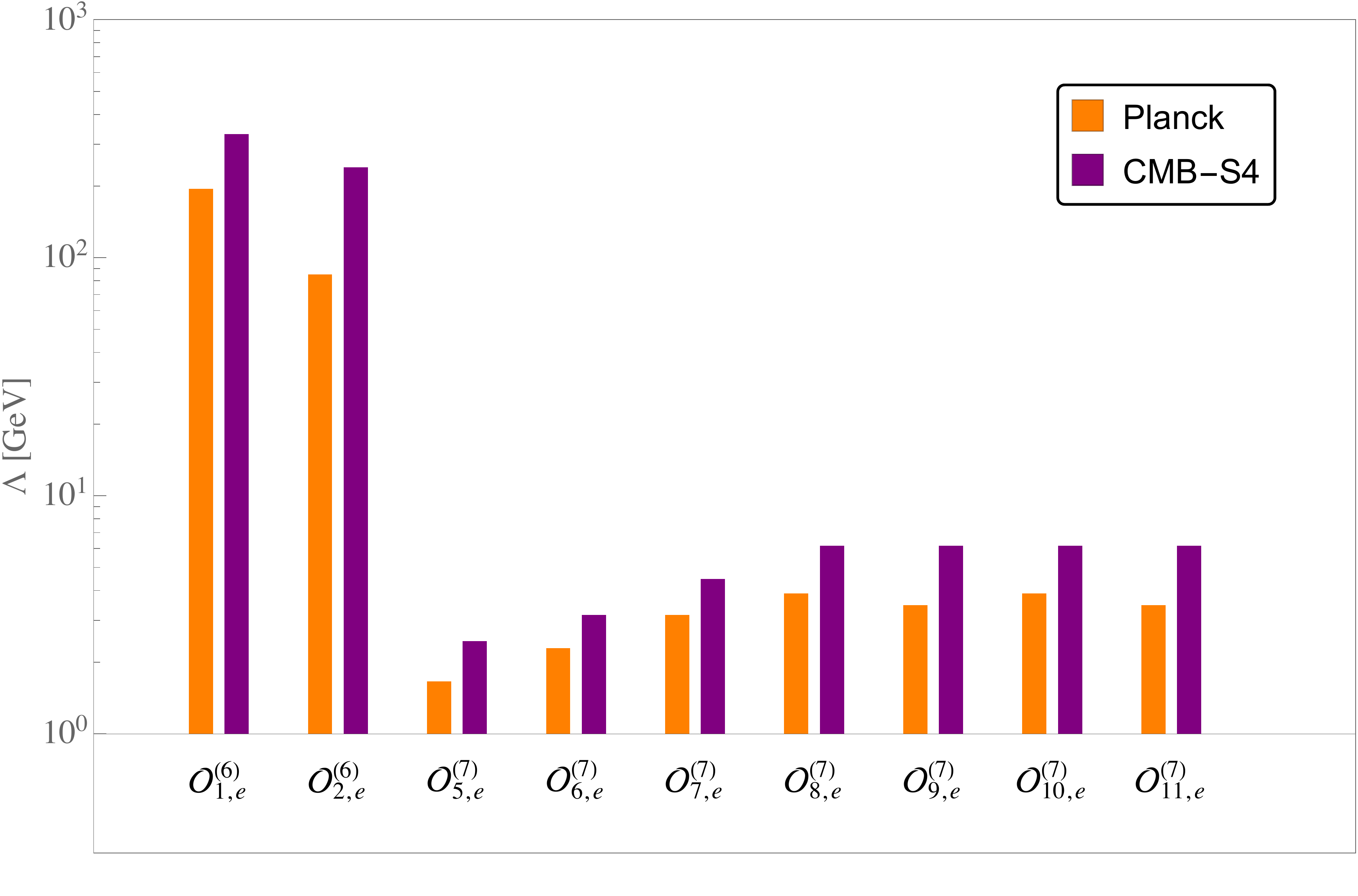}}
        \caption{Constraints on the characteristic scale $\Lambda$ of new physics from $\Delta N_{\rm eff}=N_{\rm eff}^{\rm SM+EFT}-N_{\rm eff}^{\rm SM}$, where $N_{\rm eff}^{\rm SM}=3.044$\,\cite{Akita:2020szl,Froustey:2020mcq} is the SM prediction of $N_{\rm eff}$, and $N_{\rm eff}^{\rm SM+EFT}$ is that from SM and new physics. Note that the plot is obtained by fixing all Wilson coefficients at unity and considering only one non-vanishing operator only at a time. See the main text for more discussion.}
        \label{plt:ConstraintsNP}
  \end{figure}
  
With the complete dictionary presented in section\,\ref{sec:SMEFT}, one can readily solve the Boltzmann equations for $T_\gamma$ and $T_{\nu_\alpha}$, and thus obtain corrections to $N_{\rm eff}$. In what follows, we define these corrections as 
\eqal{
\Delta N_{\rm eff} = N_{\rm eff}^{\rm SM+EFT} - N_{\rm eff}^{\rm SM},
} 
where $N_{\rm eff}^{\rm SM+EFT}$ is the theoretical prediction of $N_{\rm eff}$ with the inclusion of the NC NSI operators, and $N_{\rm eff}^{\rm SM}=3.044$\,\cite{Akita:2020szl,Froustey:2020mcq} that from the pure SM. For Planck, we use the current result $N_{\rm eff}=2.99^{+0.34}_{-0.33}$\,\cite{Aghanim:2018eyx} at the 95\% CL to obtain the constraints, and $\Delta N_{\rm eff}<0.06$ at 95\% CL for CMB-S4\,\cite{Abazajian:2016yjj,Abazajian:2019tiv,Abitbol:2017nao,Abazajian:2019eic}.

Our code is built upon {\tt nudec\_BSM} from Ref.\,\cite{Escudero:2020dfa}, and is then used to solve eqs.\,(\ref{TvFinal}-\ref{TgammaFinal}) numerically by {\tt Mathematica}. During our numerical solutions, we keep terms proportional to $m_e$ in $\langle\mathcal{M}^{2}\rangle$ and assume $T_{\nu_\mu}=T_{\nu_\tau}\ne T_{\nu_e}$ and $\mu_{\nu_\mu}=\mu_{\nu_\tau}\ne \mu_{\nu_e}$. In the very large $\Lambda$ limit, we reproduce the results in Table 1 of Ref.\,\cite{Escudero:2020dfa} for both $T_{\nu_e}=T_{\nu_{\mu,\tau}}$ and $T_{\nu_e}\ne T_{\nu_{\mu,\tau}}$. We then show our results for varying Wilson coefficients or the new physics scale $\Lambda$ in the following subsections.

\subsection{Constraints on $\Lambda$ with fixed Wilson coefficients}
Following the notations clarified in table\,\ref{SMEFTNeffOperators} and fixing the Wilson coefficients at unity, we present our results in figure\,\ref{plt:ConstraintsNP}. The constraints shown in figure\,\ref{plt:ConstraintsNP} are obtained by assuming only one non-vanishing NSI operator at a time, and the results are presented from considering the latest results from Planck\,\cite{Aghanim:2018eyx} in orange and the proposed precision goal of CMB-S4\,\cite{Abazajian:2016yjj,Abazajian:2019tiv,Abitbol:2017nao,Abazajian:2019eic} in purple. Several points from this plot merit emphasizing: 
\begin{itemize}
\item Constraints on dimension-6 EFT operators are generically stronger than those on the dimension-7 ones. The reason is that dimension-7 operators are more suppressed by one more power of $\Lambda$. Moreover, among the dimension-6 operators, currently, the Planck data leads to the most stringent constraint on the $\mathcal{O}_{1,e}^{(6)}$ operator, whose lower bound is presently constrained to be about 195\,GeV. In the future, CMB-S4 would improve this lower bound to about 240\,GeV, as one can see from the first purple histogram in figure\,\ref{plt:ConstraintsNP}. Quantitively, we summarize the lower bounds on $\Lambda$'s for all the operators shown in figure\,\ref{plt:ConstraintsNP} in table\,\ref{tab:LamLowerBound}.

\begin{table}[t]
 \renewcommand{\arraystretch}{1.2}
    \centering
    \begin{tabular}{>{\centering}p{0.2\textwidth}p{0.2\textwidth}p{0.35\textwidth}>{\centering\arraybackslash}p{0.25\textwidth}}
        \toprule
        \multirow{2}{*}{Operators} & \multicolumn{2}{c}{Lower bound on $\Lambda$ [GeV]} \\
        \cmidrule{2-3} \\
        {} & Planck & CMB-S4 \\
        \midrule
        $\mathcal{O}_{1,e}^{(6)}$ & 194.98& 331.13 \\
        $\mathcal{O}_{2,e}^{(6)}$ & 85.11 & 239.88,\,\,\,\,except (94.84, 102.33) \\
	$\mathcal{O}_{5,e}^{(7)}$ & 1.66 & 2.45,\quad\,\,\,\,except (1.91, 2.45) \\
	$\mathcal{O}_{6,e}^{(7)}$ & 2.29 & 3.16 \\
	$\mathcal{O}_{7,e}^{(7)}$ & 3.16 & 4.47 \\
	$\mathcal{O}_{8,e}^{(7)}$ & 3.89 & 6.17 \\
	$\mathcal{O}_{9,e}^{(7)}$ & 3.47 & 6.17 \\
	$\mathcal{O}_{10,e}^{(7)}$ & 3.89 & 6.17 \\
	$\mathcal{O}_{11,e}^{(7)}$ & 3.47 & 6.17 \\
        \bottomrule
    \end{tabular}\caption{Constraints on EFT operators from current Planck data and future CMB-S4 proposal. All lower bounds are obtained by assuming one non-vanishing EFT operator at a time and fixing the Wilson coefficients at unity. Note that for $\mathcal{O}_{2,e}^{(6)}$ and $\mathcal{O}_{5,f}^{(7)}$, there are exception intervals for CMB-S4 as a result of destructive interference or a negative shift in $N_{\rm eff}$ from the EFT operators. See the main text for more discussion.}\label{tab:LamLowerBound}
\end{table}

\item As one can see from table\,\ref{tab:LamLowerBound}, for the $\mathcal{O}_{2,e}^{(6)}$ and $\mathcal{O}_{5,f}^{(7)}$ operators, there exist intervals that can not be covered by future CMB-S4 if one considers only one operator at a time. However, if one considers multiple operators, these two exception intervals would be ruled out by future CMB-S4 result. For this reason, figure \,\ref{plt:ConstraintsNP} is plotted by using the lower bounds 239.88\,GeV and 2.45\,GeV for $\mathcal{O}_{2,e}^{(6)}$ and $\mathcal{O}_{5,f}^{(7)}$ respectively. On the other hand, the exception interval for $\mathcal{O}_{2,e}^{(6)}$ results from the destructive interference between the SM and the NSI operators, while that for $\mathcal{O}_{5,f}^{(7)}$ results from a negative shift to $N_{\rm eff}$ when $\Lambda$ is small. We show this point in the second row of figure\,\ref{plt:DelNeffPlots}.

\begin{figure}[t]
\centering{
  \begin{adjustbox}{max width = \textwidth}
\begin{tabular}{cc}
\includegraphics[scale=0.5]{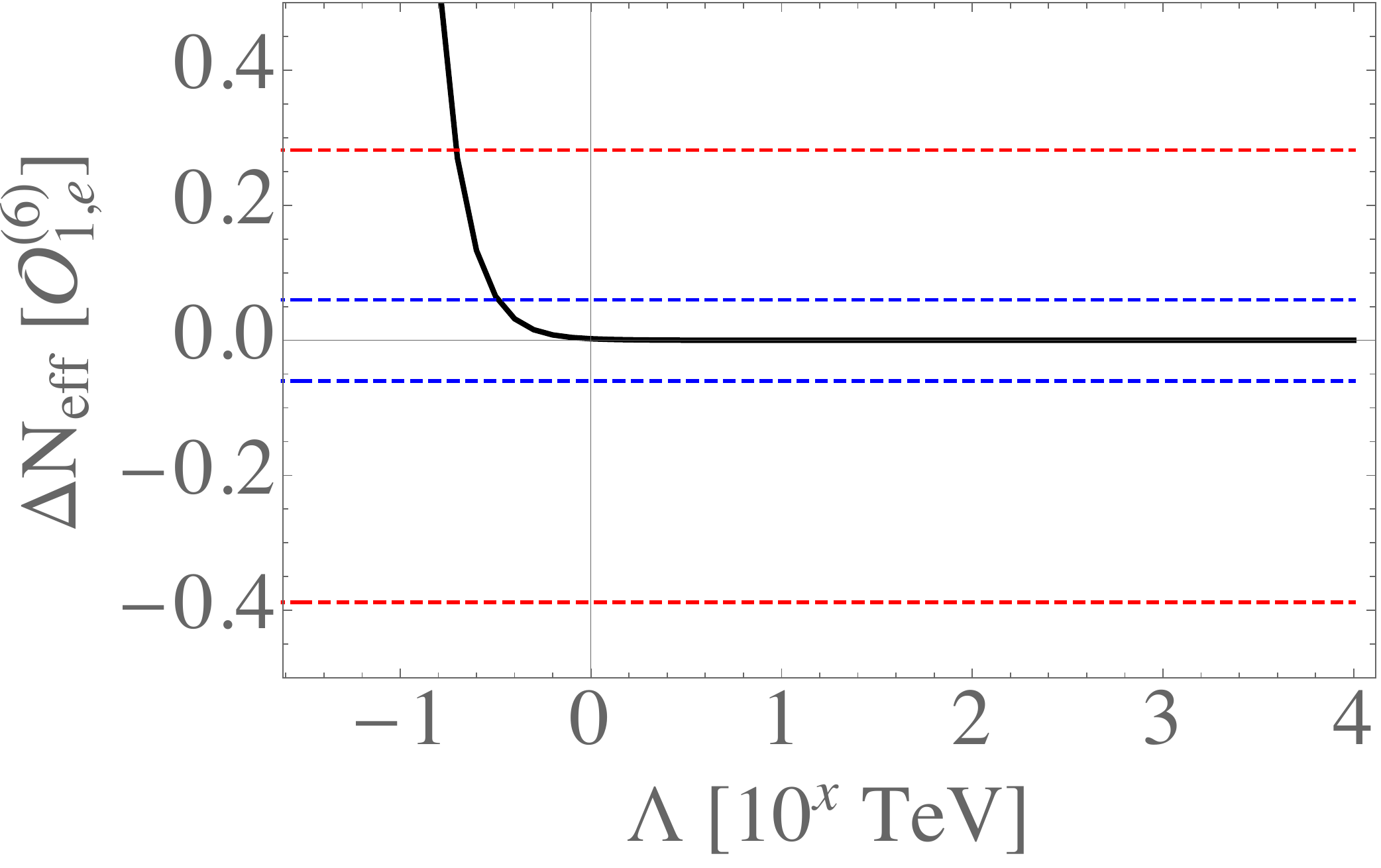} ~& ~ \includegraphics[scale=0.5]{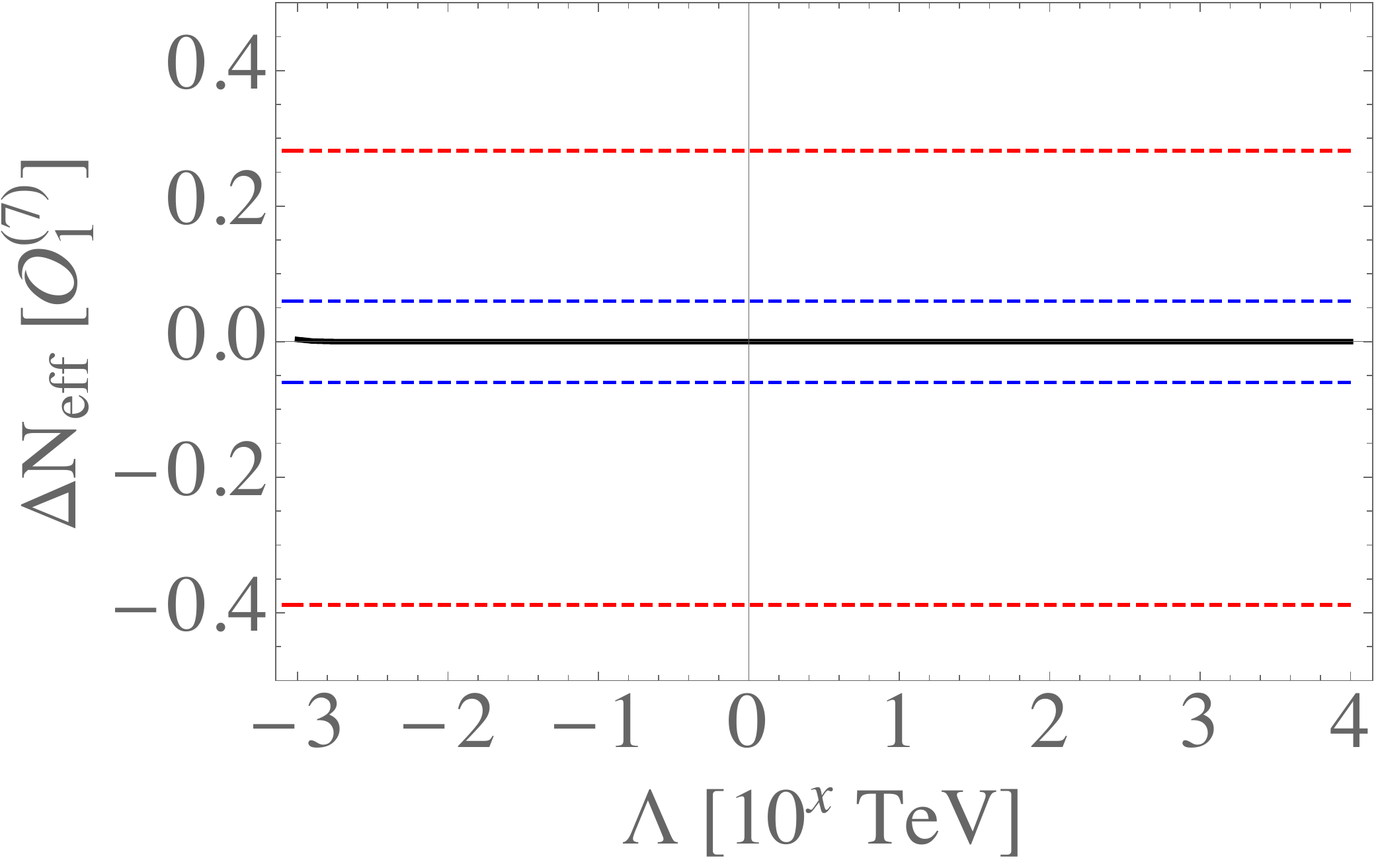}\\
\includegraphics[scale=0.5]{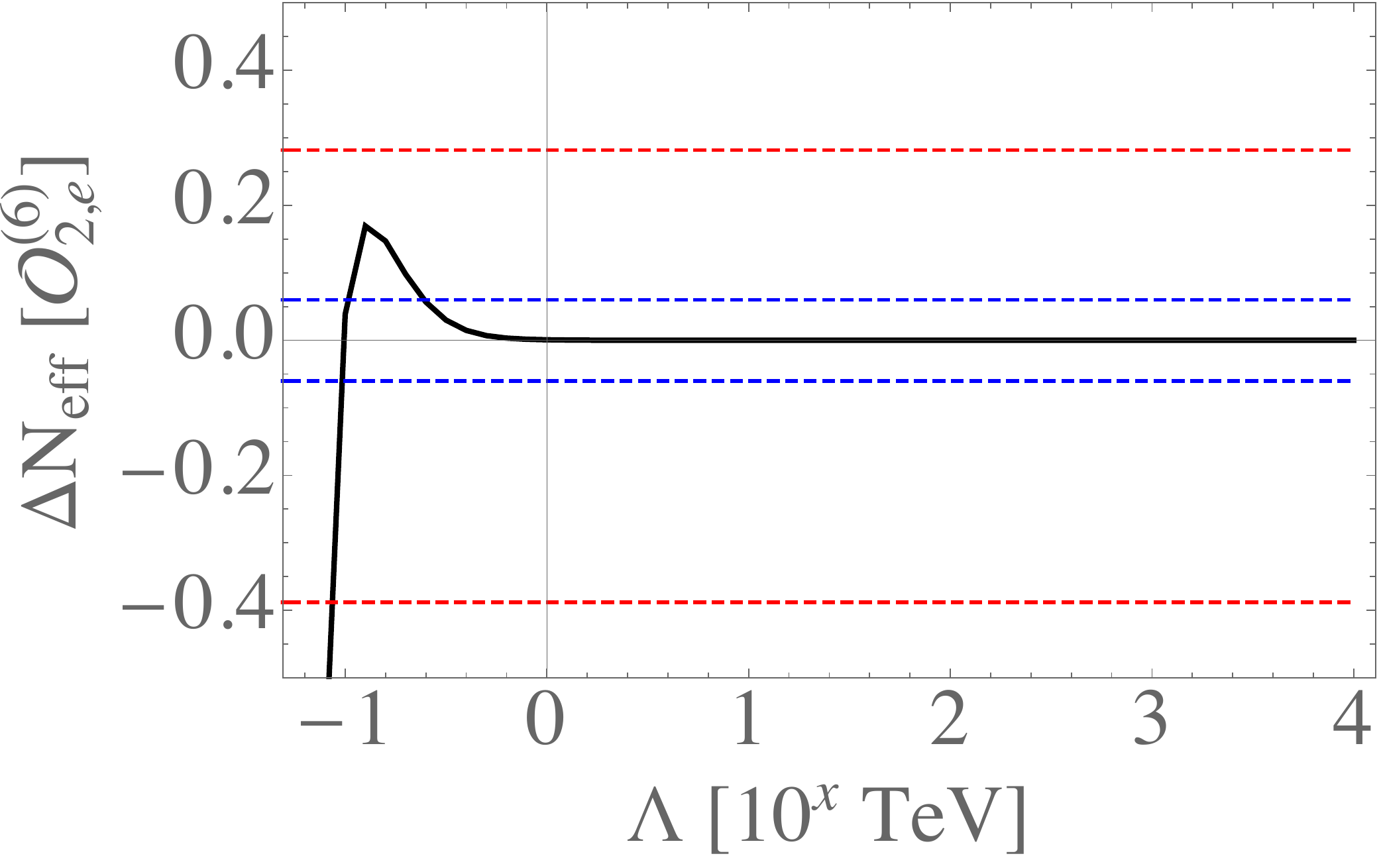} &\includegraphics[scale=0.5]{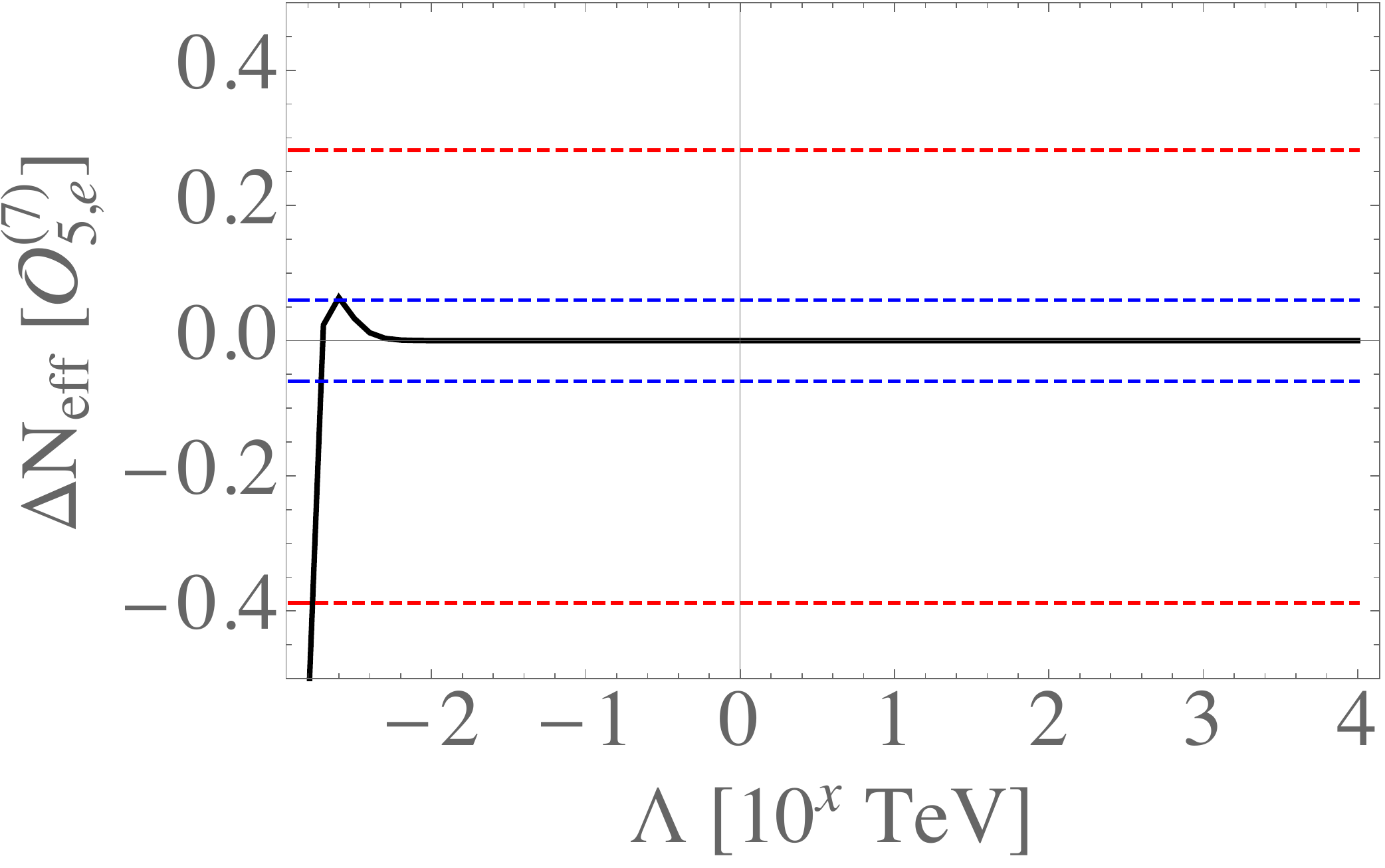}
\end{tabular}
  \end{adjustbox}}
\caption{Corrections to $N_{\rm eff}$ from varying $\Lambda$. Upper left: $\Delta N_{\rm eff}$ with the inclusion of $\mathcal{O}_{1,e}^{(6)}$ only. Very similar plots are obtained for other operators except for $\mathcal{O}_{(2,e),3,4,5}^{(6)}$, $\mathcal{O}_{1,2, (5,e)}^{(7)}$, thus we only show $\mathcal{O}_{1,e}^{(6)}$ for illustration. Upper right: $\Delta N_{\rm eff}$ from $\mathcal{O}_{1}^{(7)}$ only. Very similar plot is obtained for $\mathcal{O}_{2}^{(7)}$. Lower left: $\Delta N_{\rm eff}$ from $\mathcal{O}_{2,e}^{(6)}$ only. Lower right: $\Delta N_{\rm eff}$ from $\mathcal{O}_{5,e}^{(7)}$ only. In all these subfigures, the black curve stands for corrections to $N_{\rm eff}$ from the related NSI operator, the horizontal red dashed line is the constraint on $\Delta N_{\rm eff}$ from Planck, and the horizontal red dashed line is that from CMB-S4.}\label{plt:DelNeffPlots}
\end{figure}

\item Constraints on dimension-6 operators $\mathcal{O}_{3,4,5}^{(6)}$ are missing in figure\,\ref{plt:ConstraintsNP}. The reason can be understood as follows: (1) When  $\Lambda\gtrsim \Lambda_W$ or $\Lambda\gg \Lambda_W$, SM contributions dominate and the resulting $N_{\rm eff}$ always agrees with the SM prediction -- The deviation of $N_{\rm eff}$ from the SM prediction is always within the uncertainties of both Planck and CMB-S4; (2) For $\Lambda\ll \Lambda_W$, one might na\"ively think the $\langle \mathcal{M}_{\rm SM}^2\rangle$ term in eq.\,\eqref{amp2Inter} can be safely discarded and very large $N_{\rm eff}$ would be predicted from $\mathcal{O}_{3,4,5}^{(6)}$. However, as we already point out right below eq.\,\eqref{amp2Inter}, the SM part can not be ignored since in this case, it is the only part that governs the evolution of $T_\gamma$. Furthermore, when $\Lambda\ll\Lambda_W$, neutrino self interactions are rapid enough to eliminate any difference between $T_{\nu_e}$ and $T_{\nu_{\mu,\tau}}$, and neutrinos of all flavors have exactly the same temperature.\footnote{With our choice of the Wilson coefficients and the small $\Lambda$, neutrino self-interacting rates are always larger than the Hubble rate such that the three flavor neutrinos always stay in thermal equilibrium. However, neutrino decoupling is not affected since photon-electron-positron sector is governed by weak interactions. We emphasize that the equal temperature of neutrinos is the direct result of neutrino self-interactions introduced by $\mathcal{O}_{3,4,5}^{(6)}$, the moderate Wilson coefficients, and the small $\Lambda$.} This in turn results in vanishing corrections to the collision term integrals for $\mathcal{O}_{3,4,5}^{(6)}$ as discussed right after eq.\,\eqref{coll:ampconst}. Thus, when $\Lambda$ is very small, corrections from $\mathcal{O}_{3,4,5}^{(6)}$ to $N_{\rm eff}$ vanish and the SM prediction is restored.

\item While it is a very good approximation to neglect the temperature and the chemical differences among neutrinos when calculating $N_{\rm eff}$ within the SM framework, see Table 1 of Ref.\,\cite{Escudero:2020dfa} for example, this approximation does not stay valid any more in the case where new physics introduces only neutrino self interactions as the $\mathcal{O}_{3,4,5}^{(6)}$ operators. In this scenario, if one takes the equal neutrino temperature and the equal chemical potential approximation for all the three-flavor neutrinos, then the collision term integrals simply vanish such that the effects of this new physics can never be observed.

\item Constraints on $\mathcal{O}_{1}^{(7)}$ and $\mathcal{O}_{2}^{(7)}$ are also missing in figure\,\ref{plt:ConstraintsNP}, due to the suppression factors $\alpha/(12\pi)$ and $\alpha/(8\pi)$, respectively: At the amplitude level, both these two factors lead to suppression of $\mathcal{O}(10^{-4})$, thus the invariant amplitude $\langle \mathcal{M}^2\rangle$ is suppressed by a factor of $\mathcal{O}(10^{-8})$. Note also that there is no interference between the SM and $\mathcal{O}_{1,2}^{(7)}$. The upper right panel of figure\,\ref{plt:DelNeffPlots} shows the prediction of $N_{\rm eff}$ with the inclusion of $\mathcal{O}_{1}^{(7)}$, and similar result is obtained for $\mathcal{O}_{2}^{(7)}$.

\item Though we do not consider the magnetic dipole operator $\mathcal{O}_{1}^{(5)}$ in this work, and the $\mathcal{O}_{1,2}^{(7)}$ operators are not constrained by $N_{\rm eff}$ as discussed above, lower bounds on $\Lambda$ for these operators do exist from other experiments. For $\mathcal{O}_{1}^{(5)}$, it was concluded in Ref.\,\cite{Altmannshofer:2018xyo} that the most stringent lower bound on $\Lambda$ was $2.7\times10^6$\,GeV from the magnetic moment of $\nu_e$ using Borexino Phase-II solar neutrino data\,\cite{Borexino:2017fbd}. On the other hand, translating this constraint on the magnetic moment of $\nu_e$ from $\mathcal{O}_{1,2}^{(7)}$, they found $\Lambda>328$\,GeV and $\Lambda>1081$\,GeV for $\mathcal{O}_{1}^{(7)}$ and $\mathcal{O}_{2}^{(7)}$ respectively. Furthermore, $\mathcal{O}_{1,e}^{(6)}$ was also constrained to have a lower bound of 1005\,GeV from a global fitting of neutrino oscillating data\,\cite{Esteban:2018ppq,Altmannshofer:2018xyo}. As one can see from table\,\ref{tab:LamLowerBound}, the constraint on $\mathcal{O}_{1,e}^{(6)}$ from the global fitting is stronger than that from $N_{\rm eff}$. However, all the other operators in figure\,\ref{plt:ConstraintsNP} are not constrained in Ref.\,\cite{Altmannshofer:2018xyo}, making our work complementary to theirs as well as that in Ref.\,\cite{Du:2020dwr}.
\end{itemize}

We emphasize that conclusions above are obtained by fixing the Wilson coefficients at one and considering only one non-vanishing NSI operator at a time. In Ref.\,\cite{Du:2020dwr}, we find that if multiple operators exist at the same scale, then the correlation among them may change the constraints by orders of magnitude. However, due to the computation challenge, this correlation effect is in general ignored except for some UV models where the number of operators at certain dimension is limited. In this work, we find when $\Lambda\sim\Lambda_W$ or smaller where NSI contributions to $\langle\mathcal{M}^2\rangle$ are comparable to or dominate over those from the SM, numerical computation of the Boltzmann equations is extremely slow or even impossible even with high-performance clusters. For this reason, the correlation mentioned above will not be discussed.

\subsection{Constraints on Wilson coefficients with fixed $\Lambda$}

\begin{figure}[t]
\centering{
  \begin{adjustbox}{max width = \textwidth}
\begin{tabular}{cc}
\includegraphics[scale=0.5]{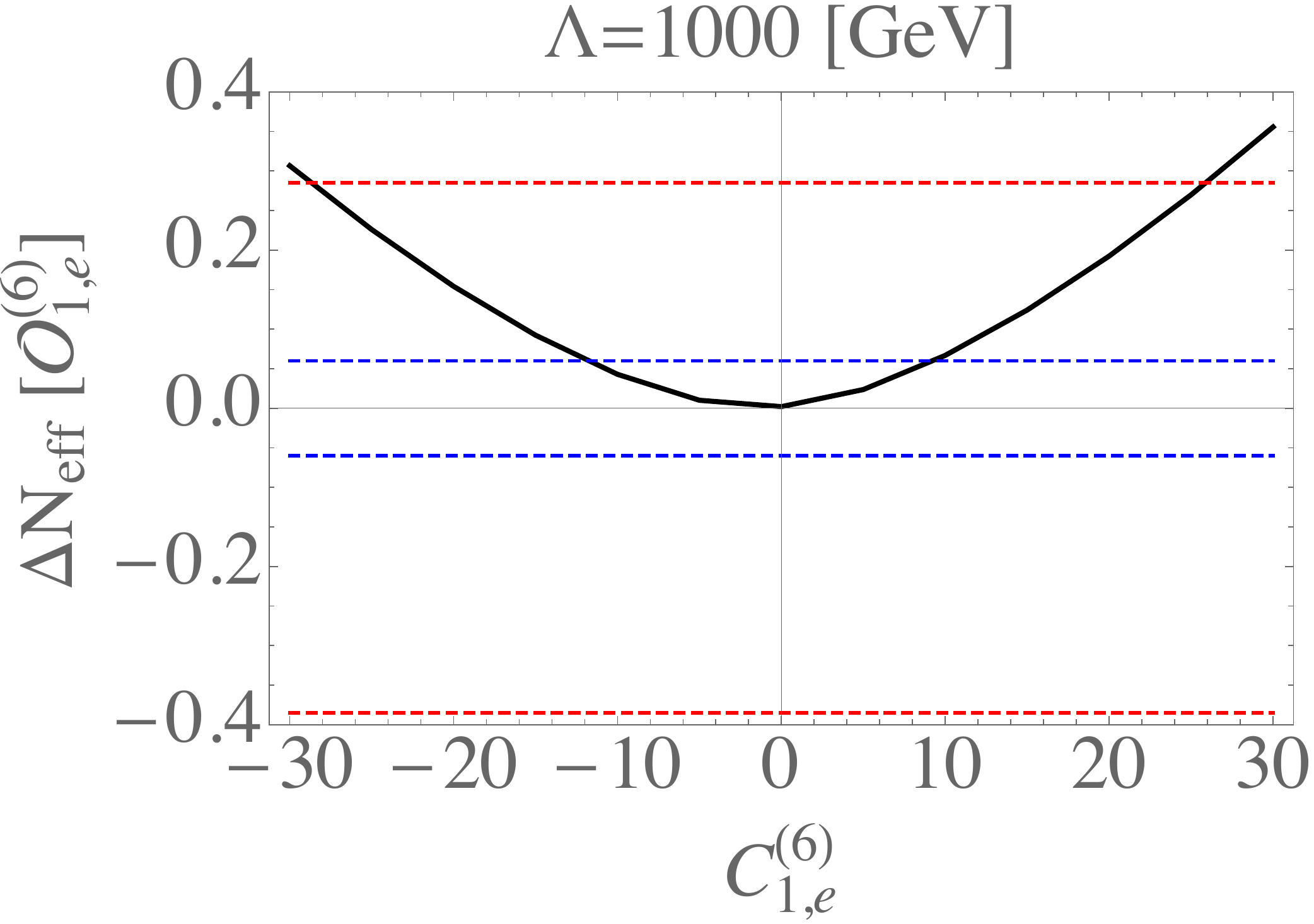} ~& ~ \includegraphics[scale=0.5]{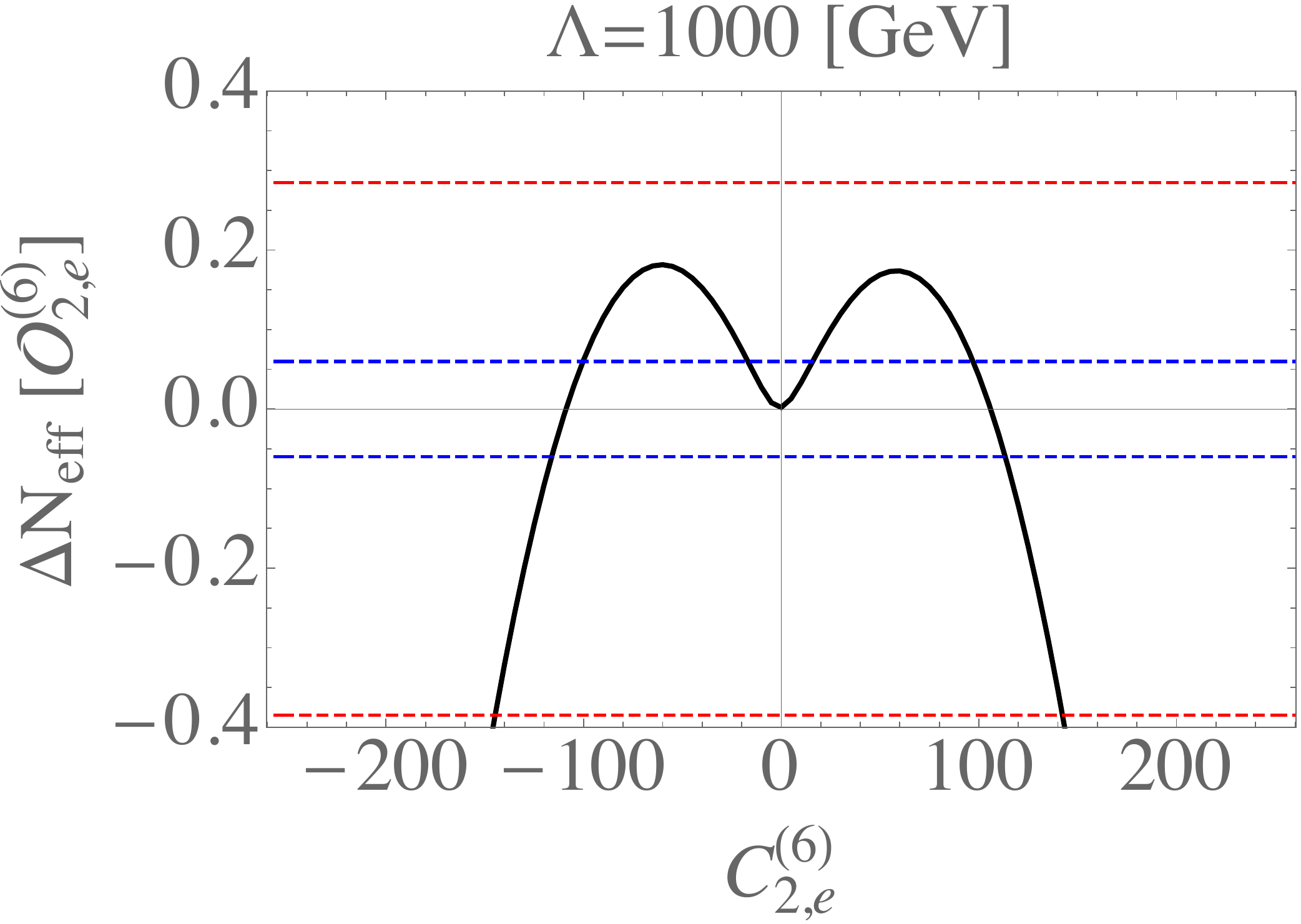}\\
\includegraphics[scale=0.5]{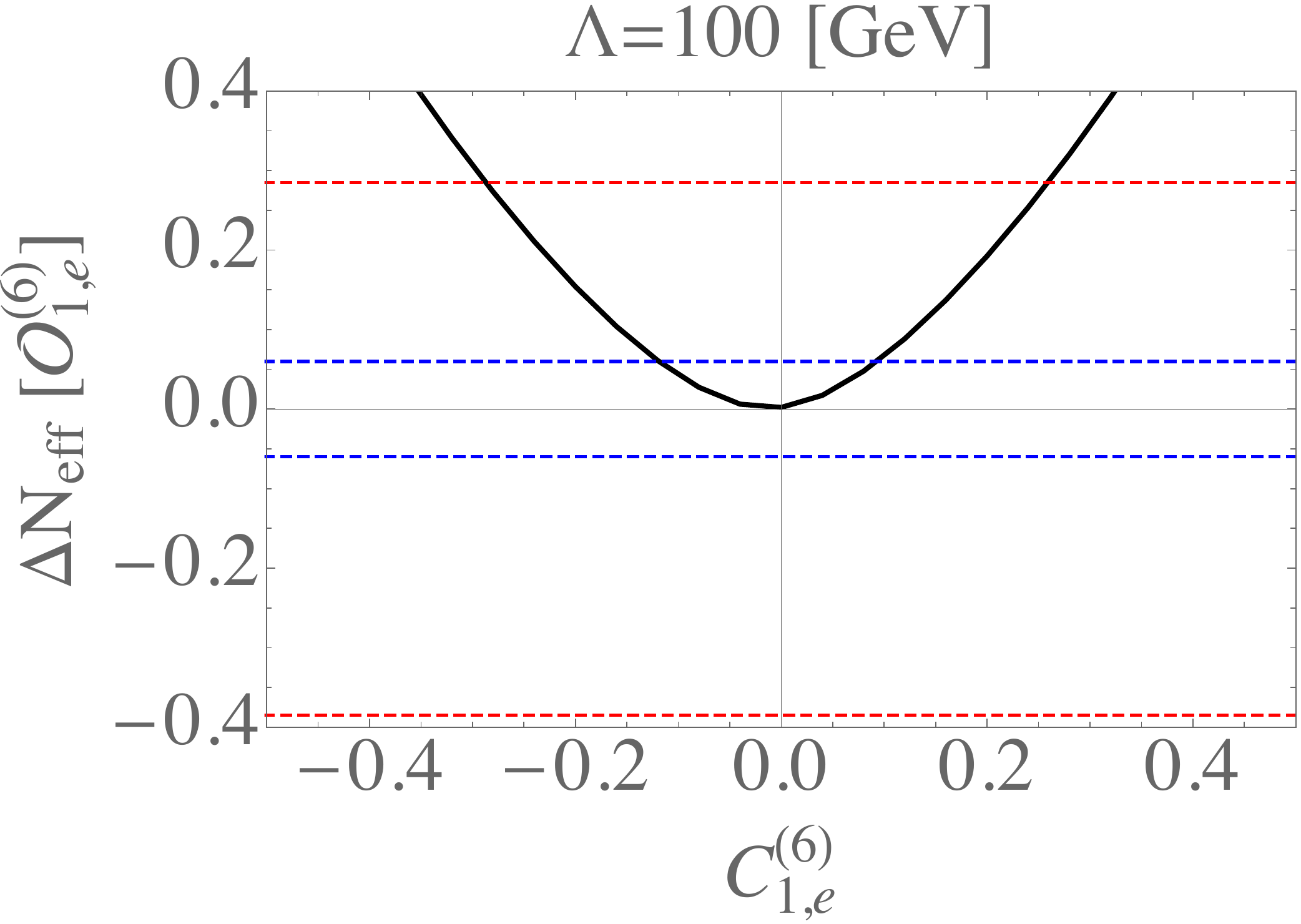} ~& ~ \includegraphics[scale=0.5]{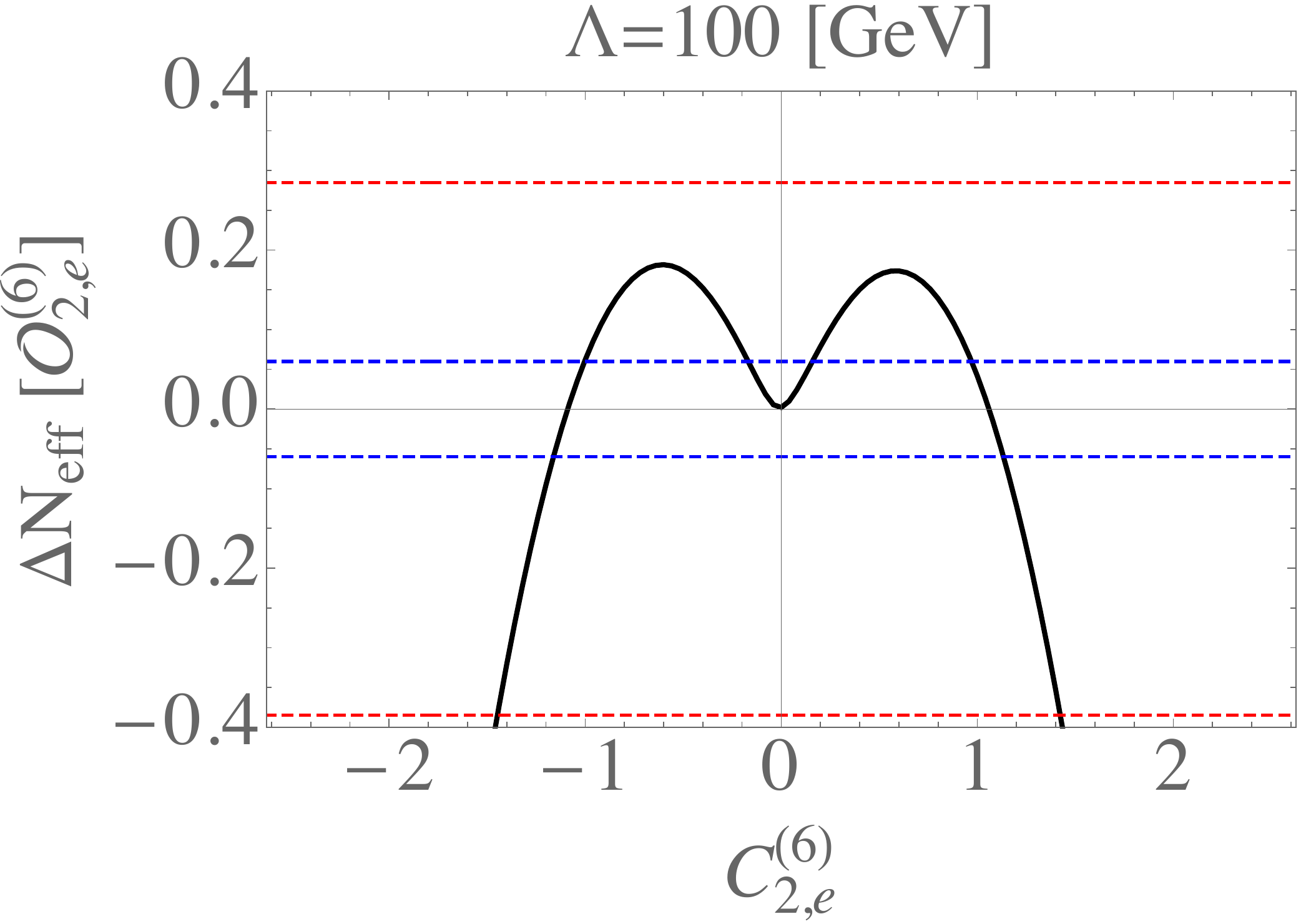}\\
\end{tabular}
  \end{adjustbox}}
\caption{Corrections to $N_{\rm eff}$ from varying Wilson coefficients with $\Lambda=1$\,TeV and by considering only one non-vanishing NSI operator at a time. The upper left (right) panel corresponds to $\Delta N_{\rm eff}$ from $\mathcal{O}_{1(2),e}^{(6)}$ with $\Lambda=1000$\,GeV, and the lower left (right) panel is the same but with $\Lambda=100$\,GeV. The black curve stands for corrections to $N_{\rm eff}$ from the NSI operator, and the horizontal colorful lines have the same meaning as those in figure\,\ref{plt:DelNeffPlots}. Note the scale difference of the horizontal axes and see more discussion in the main text.}\label{plt:DelNeffPlots2}
\end{figure}

Alternatively, we present the constraints for the Wilson coefficients with $\Lambda=1$\,TeV and 100\,GeV in this subsection, and the results are shown in the first and the second rows of figure\,\ref{plt:DelNeffPlots2}, respectively. Constraints are shown for $\mathcal{O}_{1,e}^{(6)}$ and $\mathcal{O}_{2,e}^{(6)}$ only, and all the other Wilson coefficients stay unconstrained for the range we consider in figure\,\ref{plt:DelNeffPlots2}. Quantitively, we find, assuming the same Wilson coefficients for neutrinos of different flavors,
\begin{itemize}
\item For $\Lambda=1$\,TeV:
\eqal{
-28.7\lesssim C_{1,e}^{(6)}\lesssim 25.8\quad {\rm (Planck),} &\quad -11.8\lesssim C_{1,e}^{(6)}\lesssim 9.0\quad \text{ (CMB-S4)}\label{C61WilCoeff}\\
-145.2\lesssim C_{2,e}^{(6)}\lesssim 141.3\quad {\rm (Planck),} &\quad -17.0\lesssim C_{2,e}^{(6)}\lesssim 15.8\quad \text{ (CMB-S4),}\label{C62WilCoeff}\\
\text{except for }  &C_{2,e}^{(6)}\in(-116.7, -100.7) \cup (96.4, 112.5) {\text{ for CMB-S4}.}\nonumber
}
\item For $\Lambda=100$\,GeV:
\eqal{
-0.29\lesssim C_{1,e}^{(6)}\lesssim 0.26\quad {\rm (Planck),} &\quad -0.12\lesssim C_{1,e}^{(6)}\lesssim 0.09\quad \text{ (CMB-S4)}\label{C61WilCoeff2}\\
-1.45\lesssim C_{2,e}^{(6)}\lesssim 1.42\quad {\rm (Planck),} &\quad -0.18\lesssim C_{2,e}^{(6)}\lesssim 0.15\quad \text{ (CMB-S4),}\label{C62WilCoeff2}\\
\text{except for }  &C_{2,e}^{(6)}\in(-1.17, -1.01) \cup (0.96, 1.13) {\text{ for CMB-S4}.}\nonumber
}
\end{itemize}
For $C_{2,e}^{(6)}$, the two exception intervals for CMB-S4 in the last line of the two bullets above result from destructive interference as already discussed in last subsection -- For $C_{2,e}^{(6)}$ of $\mathcal{O}(10)$ or larger, $\mathcal{O}_{1,e}^{(6)}$ is effectively of the weak scale, leading to the destructive interference and thus the two intervals. {\color{black}This can be understood more explicitly from the analytical expressions of the neutrino total energy densities from $\mathcal{O}_{1,e}^{(6)}$ and $\mathcal{O}_{2,e}^{(6)}$:\footnote{{\color{black}These results can be readily obtained by using our complete dictionary in section\,\ref{sec:ComDictColl} or the analytical expressions in the auxiliary {\tt Mathematica} notebook file.}}
\eqal{
\rho^{\rm interf.}_{\nu-\rm total}(\mathcal{O}_{1, e}^{(6)})\simeq&+\frac{256\sqrt{2}C_{1,e}^{(6)}G_F\sin^2\theta_WT_\gamma^9}{\pi^5\Lambda^2},\\
\rho^{\rm interf.}_{\nu-\rm total}(\mathcal{O}_{2, e}^{(6)})\simeq&-\frac{40\sqrt{2}C_{2,e}^{(6)}G_FT_\gamma^5T_{\nu_e}^4}{\pi^5\Lambda^2}\times(1+4\sin^2\theta_W),
}
where $\theta_W$ is the weak mixing angle and we only show the interfering terms here by omitting any sub-leading effects in them in each case. Note that a larger (smaller) neutrino energy density would be equivalent to a higher (lower) neutrino temperature. Thus, as can be understood from eq.\eqref{eq:nefffinal}, the constructive (destructive) interference also explains the positive (negative) shift feature of $N_{\rm eff}$ from $\mathcal{O}_{1,e}^{(6)}$ ($\mathcal{O}_{2,e}^{(6)}$) in figures\,\ref{plt:DelNeffPlots} and \ref{plt:DelNeffPlots2} when $\Lambda\sim\Lambda_W$.} For the other NSI operators not shown in figure\,\ref{plt:DelNeffPlots2}, since they are at least suppressed by one more power of $\Lambda$, Planck and CMB-S4 are not able to constrain those Wilson coefficients when $\Lambda$ is fixed at 1\,TeV. Similar observation is obtained for $\Lambda=100$\,GeV, with stronger constraints on $C_{1,e}^{(6)}$ and $C_{2,e}^{(6)}$, whose magnitudes are 100 times smaller compared with the $\Lambda=1$\,TeV case as expected.

\subsection{Comparison with current constraints on NC NSIs}

\begin{figure}[t]
\centering{
  \begin{adjustbox}{max width = \textwidth}
\begin{tabular}{cc}
\includegraphics[scale=0.5]{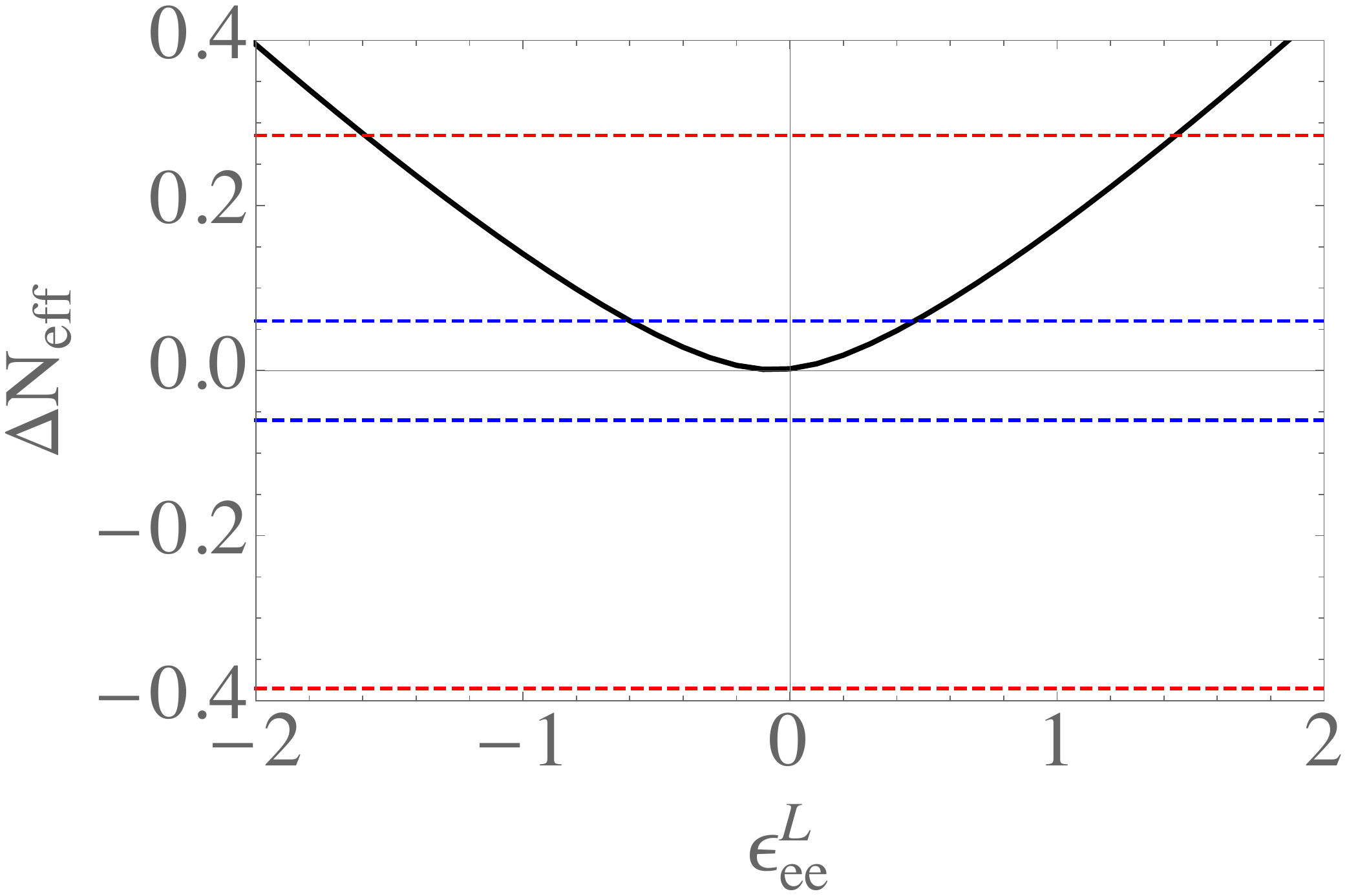} ~& ~ \includegraphics[scale=0.5]{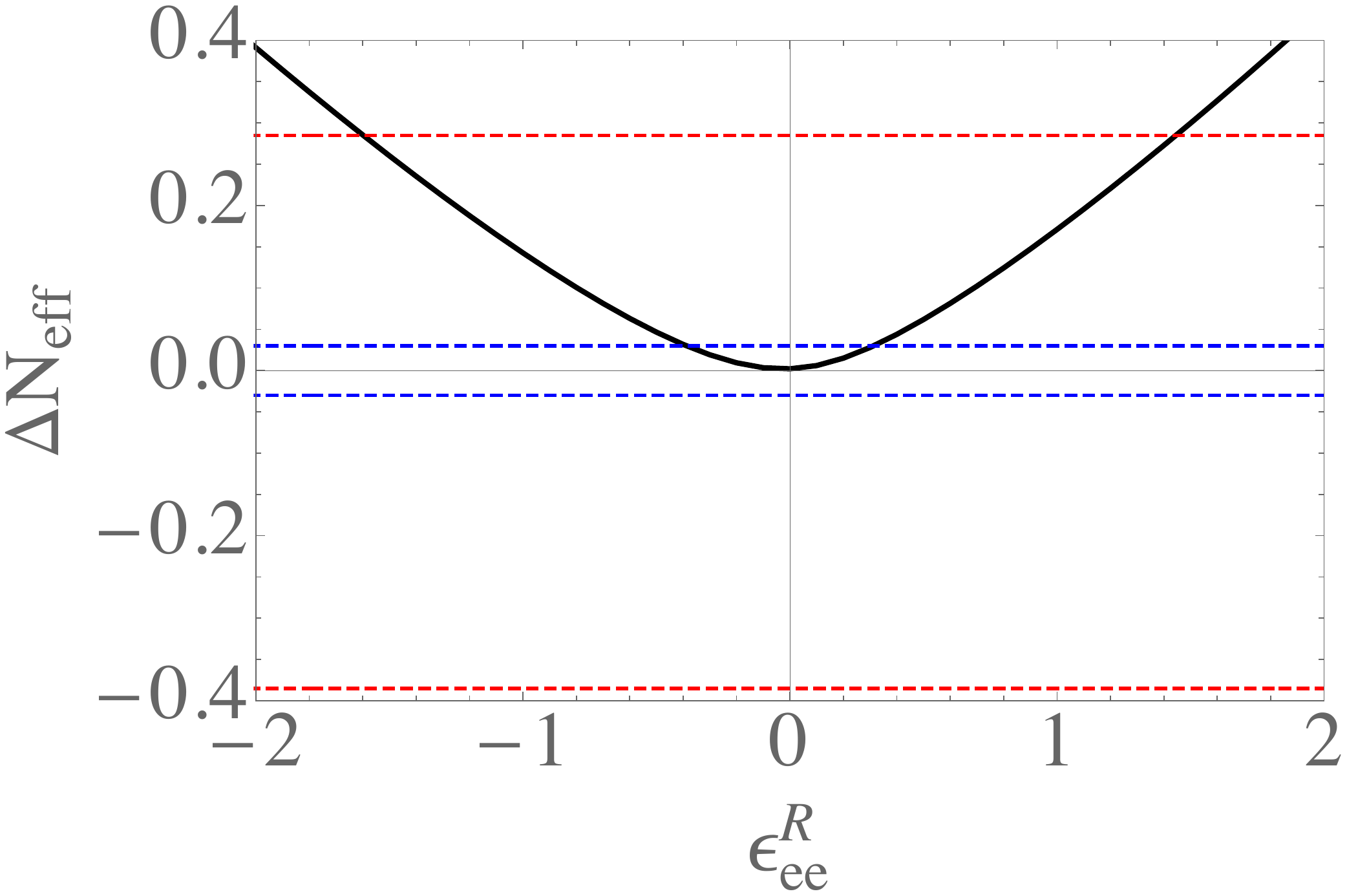}
\end{tabular}
  \end{adjustbox}}
\caption{Constraints on $\epsilon_{e,L}$ (left panel) and $\epsilon_{e,R}$ (right panel) from $N_{\rm eff}$. The black curves stand for corrections to $N_{\rm eff}$ from the dimension-6 NC NSI operators, and the colorful lines are the same as those in figure\,\ref{plt:DelNeffPlots}. See more details on the notations used in these two plots.}\label{plt:eLeRconstr}
\end{figure}

To compare with constraints on the dimension-6 operators $\mathcal{O}_{1(2),e}^{(6)}$ from other experiments, we first review the parameterization commonly used in the literatures to describe neutrino NSIs:
\eqal{
\mathcal{L}_{\rm NSI}^{\rm NC}=-2\sqrt{2}G_F\sum_{\alpha,\beta,f,P}\epsilon_{\alpha\beta}^{f,P}\left(\bar{\nu}_\alpha\gamma_\mu P_L\nu_\beta\right)\left( \bar{f}\gamma^\mu P f \right)\label{lag:ncnsi},
}
with $f=e,u,d$ the charged fermioins, $\alpha,\beta=e,\mu,\tau$ the flavor of neutrinos, and $P=L,R$ the chiral projector operators where $L=(1-\gamma_5)/2$ and $R=(1+\gamma_5)/2$. Note that the nine $\epsilon_{\alpha\beta}^{f,P}$'s are all real, and Hermiticity of the Lagrangian guarantees that only six of them are independent. The relevant operators for our study in this work are $\epsilon_{\alpha\beta}^{e,P}$. One readily finds, in terms of the Wilson coefficients used in this work, the $\epsilon$ parameters can be expressed as
\eqal{
\epsilon_{\alpha\beta}^{e,L} = \frac{C_{1,e}^{(6)} - C_{2,e}^{(6)}}{\Lambda^2\cdot 2\sqrt{2}G_F},\quad\epsilon_{\alpha\beta}^{e,R} = \frac{C_{1,e}^{(6)} + C_{2,e}^{(6)}}{\Lambda^2\cdot 2\sqrt{2}G_F}.\label{eq:ep2wil}
}

Fixing $\Lambda\simeq174.10$\,GeV from the $\Lambda^2\cdot 2\sqrt{2}G_F=1$ condition such that the LEFT in our notation mimics that in eq.\,\eqref{lag:ncnsi}, we present our results for $\epsilon_{\alpha\beta}^{L,R}$ in figure\,\ref{plt:eLeRconstr} by including all NC NSIs in eq.\,\eqref{lag:ncnsi} while ignoring all neutrino flavor dependence of $C_{(1,2),e}^{(6)}$. The colors in each subgraph of figure\,\ref{plt:eLeRconstr} have exactly the same meaning as those in figure\,\ref{plt:DelNeffPlots}, and to obtain the constraints, we once again ignore the neutrino flavor dependence of the LEFT Wilson coefficients and consider only one non-vanishing $\epsilon$ at a time in our analysis. However, we stress that, as one can see directly from eq.\,\eqref{eq:ep2wil}, one non-vanishing $\epsilon$ in general includes contributions from both $\mathcal{O}_{1,e}^{(6)}$ and $\mathcal{O}_{2,e}^{(6)}$.  To summarize, we find the $\epsilon$'s are constrained by $N_{\rm eff}$ as\footnote{Since we ignore the neutrino flavor dependence, we thus leave out the neutrino flavor indices here and in the following.}
\eqal{
-1.60\lesssim\epsilon^{e,L}\lesssim1.44\quad\text{(Planck)},\quad -0.61\lesssim\epsilon^{e,L}\lesssim0.46\quad\text{(CMB-S4)};\\
-1.60\lesssim\epsilon^{e,R}\lesssim1.44\quad\text{(Planck)},\quad -0.39\lesssim\epsilon^{e,R}\lesssim0.31\quad\text{(CMB-S4)}.
}

\begin{table}[t]
\centering{
  \begin{adjustbox}{max width = \textwidth}
 \renewcommand{\arraystretch}{1}
    \begin{tabular}{>{\centering}p{0.04\textwidth}p{0.16\textwidth}p{0.13\textwidth}p{0.13\textwidth}p{0.13\textwidth}p{0.16\textwidth}p{0.16\textwidth}p{0.16\textwidth}p{0.13\textwidth}p{0.16\textwidth}>{\centering}p{0.24\textwidth}>{\centering\arraybackslash}p{0.12\textwidth}p{0.12\textwidth}}
        \toprule
	\multirow{2}{*}{$\epsilon$'s} & \multirow{2}{*}{\cite{Esteban:2018ppq}} & \multirow{2}{*}{\cite{Deniz:2010mp}} & \multirow{2}{*}{\cite{Davidson:2003ha}} & \multirow{2}{*}{\cite{Barranco:2005ps}} & \multirow{2}{*}{\cite{Barranco:2007ej}} & \multirow{2}{*}{\cite{Bolanos:2008km}} & \multirow{2}{*}{\cite{Khan:2017oxw}}  &  \multirow{2}{*}{\cite{Khan:2016uon}}  & \multirow{2}{*}{\cite{Babu:2019mfe}} & \multicolumn{2}{c}{This work} \\
	\cmidrule{11-12} \\
	{} & {} & {}  & {}  & {}  & {}  & {} & {}  & {}  & {} &  Planck & CMB-S4 \\
        \midrule
        $\epsilon^{e,L}_{ee}$ & [-0.010, 2.039] & [-1.53, 0.38]  & [-0.07, 0.1] & [-0.05, 0.12]  & [-0.03, 0.08]  & [-0.036, 0.063]   & [-0.017, 0.027] [-0.003, 0.003] & [-0.08, 0.08]  & [-0.185, 0.380] [-0.130, 0.185] & \multirow{1}{*}{[-1.6, 1.44]} & \multirow{1}{*}{[-0.61, 0.46]}   \\
        \hline
        $\epsilon^{e,L}_{e\mu}$ & [-0.179, 0.146] & [-0.84, 0.84] & -  & -  & [-0.13, 0.13]  & -   & [-0.152, 0.152] [-0.055,0.055] & [-0.33, 0.35]  & [-0.025, 0.052] [-0.017, 0.040]  & \multirow{1}{*}{[-1.6, 1.44]} & \multirow{1}{*}{[-0.61, 0.46]}   \\
        \hline
        $\epsilon^{e,L}_{e\tau}$ &  [-0.860, 0.350] & [-0.84, 0.84] & [-0.4, 0.4]  & [-0.44, 0.44]  & [-0.33, 0.33]  & -    & [-0.152, 0.152] [-0.055,0.055] & [-0.33, 0.35] & [-0.055, 0.023] [-0.042, 0.012]  & \multirow{1}{*}{[-1.6, 1.44]} & \multirow{1}{*}{[-0.61, 0.46]}   \\
        \hline
	$\epsilon^{e,L}_{\mu\mu}$ & [-0.364, 1.387] & -  & [-0.03,0.03]  & -  & [-0.03, 0.03]  & {-}   & [-0.040, 0.04 ] [-0.010,0.010]  & -  & [-0.290, 0.390] [-0.192, 0.240] & \multirow{1}{*}{[-1.6, 1.44]} & \multirow{1}{*}{[-0.61, 0.46]}   \\
	\hline
	$\epsilon^{e,L}_{\mu\tau}$ & [-0.035, 0.028] & -  &  [-0.1,0.1]  & -  & [-0.1, 0.1]  & {-}   & - & -  & [-0.015, 0.013] [-0.010, 0.010] & \multirow{1}{*}{[-1.6, 1.44]} & \multirow{1}{*}{[-0.61, 0.46]}   \\
	\hline
	$\epsilon^{e,L}_{\tau\tau}$ & [-0.350, 1.400]  & - &  [-0.5,0.5]  & -  & [-0.46, 0.24]  & [-0.16 , 0.110 ] [0.41, 0.66] & [-0.040, 0.04 ] [-0.010,0.010]  & -  & [-0.360, 0.145] [-0.120, 0.095]  & \multirow{1}{*}{[-1.6, 1.44]} & \multirow{1}{*}{[-0.61, 0.46]}   \\
        \midrule
        \midrule
        $\epsilon^{e,R}_{ee}$ & [-0.010, 2.039] & [-0.07, 0.08]  & [-1, 0.5]  & [-0.04, 0.14]  & [0.004, 0.151]  & [-0.27, 0.59]   & [  -0.33 , 0.25  ] [-0.07, 0.07] & [-0.04, 0.06]  & [-0.185, 0.380] [-0.130, 0.185] & \multirow{1}{*}{[-1.6, 1.44]}   &   \multirow{1}{*}{[-0.39, 0.31]}\\
        \hline
        $\epsilon^{e,R}_{e\mu}$ &  [-0.179, 0.146] &  [-0.19, 0.19] & -  & -  & [-0.13, 0.13]  & {-}   & [-0.236, 0.236] [-0.08, 0.08] & [-0.15, 0.16]  &  [-0.025, 0.052] [-0.017, 0.040] & \multirow{1}{*}{[-1.6, 1.44]} &   \multirow{1}{*}{[-0.39, 0.31]}\\
        \hline
        $\epsilon^{e,R}_{e\tau}$ & [-0.860, 0.350] & [-0.19, 0.19] & [-0.7, 0.7]  & [-0.27, 0.27]  & [\,\,-0.05 ,  0.05\,\,]  [-0.28, 0.28]  & {-}   & [-0.236, 0.236]  [-0.08, 0.08] & [-0.15, 0.16]  & [-0.055, 0.023] [-0.042, 0.012]  & \multirow{1}{*}{[-1.6, 1.44]} &   \multirow{1}{*}{[-0.39, 0.31]}\\
        \hline
        $\epsilon^{e,R}_{\mu\mu}$ & [-0.364, 1.387] & - &  [-0.03,0.03]  & -  & [-0.03, 0.03]  & {-}   & [ -0.10 ,  0.12 ] [-0.006, 0.006]  & -  &  [-0.290, 0.390] [-0.192, 0.240]  & \multirow{1}{*}{[-1.6, 1.44]} &   \multirow{1}{*}{[-0.39, 0.31]}\\
        \hline
        $\epsilon^{e,R}_{\mu\tau}$ & [-0.035, 0.028]  & - &  [-0.1,0.1]  & -  & [-0.1, 0.1]  & {-}   & - & -  &  [-0.015, 0.013] [-0.010, 0.010] & \multirow{1}{*}{[-1.6, 1.44]} &   \multirow{1}{*}{[-0.39, 0.31]}\\
        \hline
        $\epsilon^{e,R}_{\tau\tau}$ & [-0.350, 1.400] & - & [-0.5,0.5]  & -  & [-0.25, 0.43]  & {[-1.05, 0.31]}   &  [ -0.10 ,  0.12 ] [-0.006, 0.006]   & -  & [-0.360, 0.145] [-0.120, 0.095]  & \multirow{1}{*}{[-1.6, 1.44]} &   \multirow{1}{*}{[-0.39, 0.31]}\\
        \bottomrule
    \end{tabular}
\end{adjustbox}}\caption{Summary of constraints on dimension-6 neutrino-electron NC NSIs from previous studies and this work. Constraints  from a global fitting of all kinds of neutrino oscillation data plus the COHERENT result are obtained in Ref.\,\cite{Esteban:2018ppq}, the TEXONO collaboration in Ref.\,\cite{Deniz:2010mp}, the LEP, LSND and CHARM-II experiments in Ref.\,\cite{Davidson:2003ha}, a global analysis of $\nu_ee$ and $\bar{\nu}_ee$ scattering data from LSND, Irvine, Rovno and MUNU experiments in Ref.\,\cite{Barranco:2005ps}, OPAL, ALEPH, L3, DELPHI, LSND, CHARM-II, Irvine, Rovno and MUNU experiments in Ref.\,\cite{Barranco:2007ej}, solar and reactor neutrino experiments in Ref.\,\cite{Bolanos:2008km}, low-energy solar neutrinos at source and detector from the Borexino experiment in Ref.\,\cite{Khan:2017oxw}, a global analysis of short baseline $\nu e$ and $\bar{\nu}e$ data from LSND, LAMPF, Irvine, Rovno, MUNU, TEXONO and KRANOYARSK in Ref.\,\cite{Khan:2016uon}, and DUNE in Ref.\,\cite{Babu:2019mfe}.}\label{tab:epsbounds}
\end{table}

In comparison, we list constraints on these $\epsilon$ parameters from previous studies and ours obtained in this work in table\,\ref{tab:epsbounds} by ignoring bounds from loops\,\cite{Davidson:2003ha,Biggio:2009kv}. Note that constraints from Ref.\,\cite{Esteban:2018ppq} in the second column are originally presented in terms of $\epsilon_{\alpha\beta}^{e, L+R}\equiv\epsilon_{\alpha\beta}^{e,L}+\epsilon_{\alpha\beta}^{e,R}$. We translate them on individual $\epsilon_{\alpha\beta}^{e, L(R)}$ by assuming only one of them non-vanishing.  Constraints from TEXONO are obtained at the Kuo-Sheng Nuclear Power
Station in Ref.\,\cite{Deniz:2010mp}, the LEP, LSND and CHARM-II experiments in Ref.\,\cite{Davidson:2003ha}, a global analysis of $\nu_ee$ and $\bar{\nu}_ee$ scattering data from LSND, Irvine, Rovno and MUNU experiments in Ref.\,\cite{Barranco:2005ps}, a combination of OPAL, ALEPH, L3, DELPHI, LSND, CHARM-II, Irvine, Rovno and MUNU experiments in Ref.\,\cite{Barranco:2007ej}, solar and reactor neutrino experiments in Ref.\,\cite{Bolanos:2008km}, low-energy solar neutrinos at source and detector from the Borexino experiment in Ref.\,\cite{Khan:2017oxw}, a global analysis of short baseline $\nu e$ and $\bar{\nu}e$ data from LSND, LAMPF, Irvine, Rovno, MUNU, TEXONO and KRANOYARSK in Ref.\,\cite{Khan:2016uon}, and the DUNE experiment in Ref.\,\cite{Babu:2019mfe}. For constraints from Ref.\,\cite{Barranco:2005ps}, we cite their results in the one-parameter case since it leads to the most stringent constraints on these NC NSIs, and similarly for results in Refs.\,\cite{Khan:2017oxw,Khan:2016uon}. For constraints from Ref.\,\cite{Babu:2019mfe}, the upper and the lower intervals are obtained using an exposure of 300 and 850\,kt.MW.yr for DUNE respectively. For constraints from Ref.\,\cite{Khan:2017oxw}, the upper number is obtained from a detector-only study using low-energy solar neutrinos at Borexino, while the lower is the future prospect from a combined analysis of the detector and the source. For all the other cases in table\,\ref{tab:epsbounds}, whenever two intervals appear, it means two ``disjoint'' ranges that are simultaneously allowed from their analyses. We refer the reader to the original references for more details.

As one can see from table\,\ref{tab:epsbounds}, in general, constraints from other experiments are stronger than those we obtain from Planck. However, from the last column of table\,\ref{tab:epsbounds}, the results from CMB-S4 would be improved by a factor of $\sim3$ (5) for $\epsilon^{e, L(R)}$. As a result, all the $\epsilon$'s would be bounded at the 10\% level. On the other hand, in table\,\ref{tab:epsbounds}, one notes that seven of these $\epsilon$'s are constrained at the 10\% level from previous experiments, except the following five's: $\epsilon_{ee}^{e,L}$\,\cite{Barranco:2007ej,Bolanos:2008km}, $\epsilon_{\mu\mu}^{e,(L,R)}$\,\cite{Davidson:2003ha,Barranco:2007ej}, $\epsilon_{\mu\tau}^{e,L}$\,\cite{Esteban:2018ppq}, $\epsilon_{ee}^{e,R}$\,\cite{Deniz:2010mp}, and $\epsilon_{\mu\tau}^{e,R}$\,\cite{Esteban:2018ppq} that are stringently constrained at the 1\% level. Therefore, constraints on most of these $\epsilon$'s from CMB-S4 are basically comparable to the existing ones. For example, $\epsilon_{\tau\tau}^{e,R}$ is constrained to be [-0.25, 0.43] in Ref.\,\cite{Barranco:2007ej} from the OPAL, ALEPH, L3, DELPHI, LSND, CHARM-II, Irvine, Rovno and MUNU experiments, while it would be [-0.39, 0.31] from CMB-S4. Furthermore, we point out that, in the four-parameter cases of Ref.\,\cite{Barranco:2005ps}, our results for $\epsilon^{e,R}$ from CMB-S4 are slightly stronger than theirs. Similarly, in the two-parameter (correlated) case, our constraints on $\epsilon^{e,R}_{\mu\mu,ee}$ ($\epsilon^{e,L}_{e\mu, e\tau}$) are stronger than those obtained in Ref.\,\cite{Khan:2017oxw} (\cite{Khan:2016uon}), while weaker or comparable to theirs for the other $\epsilon$'s.

\begin{figure}[t]
\centering{
  \begin{adjustbox}{max width = \textwidth}
\begin{tabular}{c}
\includegraphics[scale=0.25]{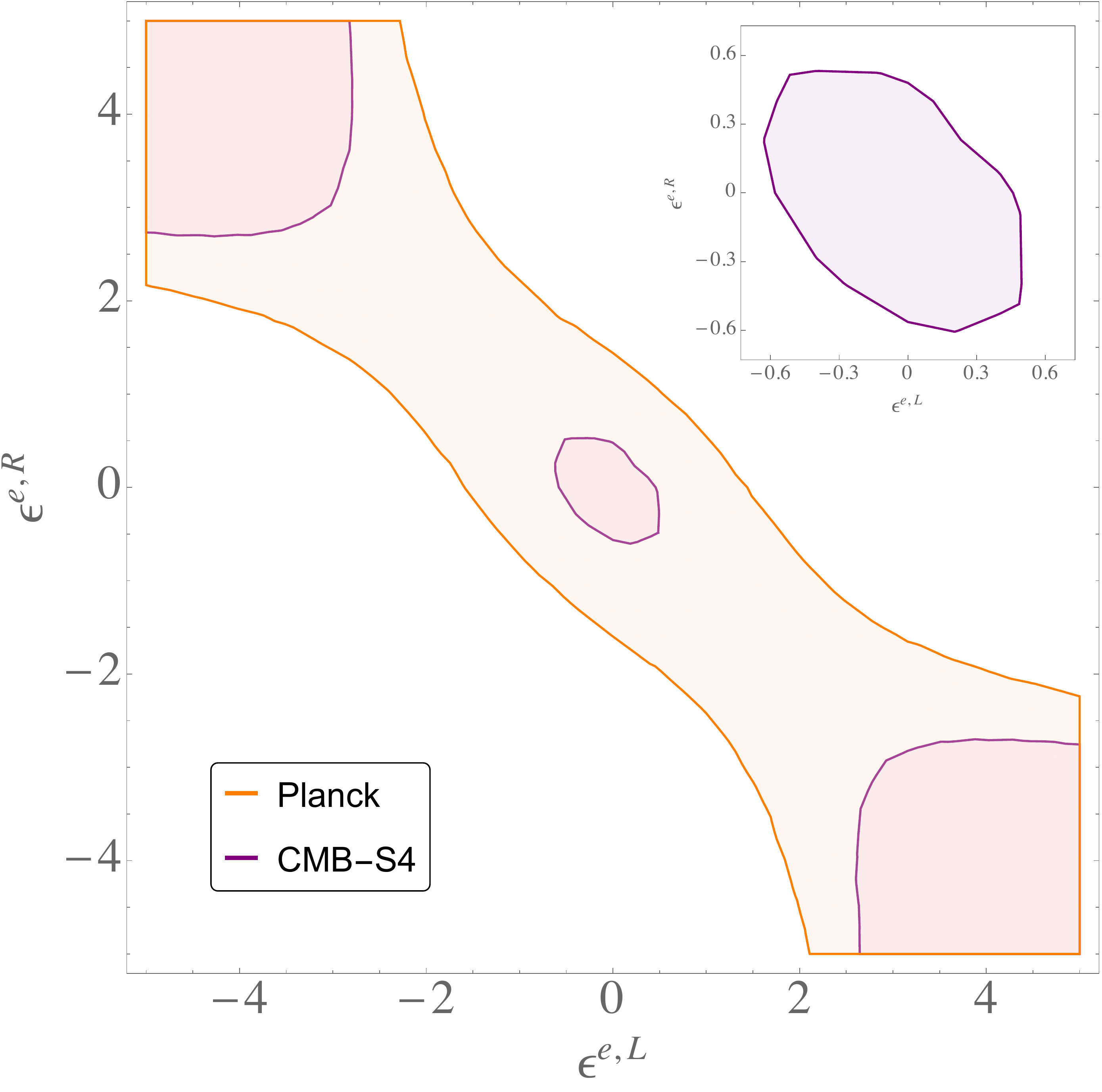}
\end{tabular}
  \end{adjustbox}}
\caption{Simultaneous constraints on $\epsilon_{e,L}$ and $\epsilon_{e,R}$ from precision measurements of $N_{\rm eff}$ from Planck and CMB-S4. The allowed regions are indicated by the orange and the purple respectively. The subgraph in the upper right corner corresponds to the magnified allowed region from CMB-S4 near the origin. See the main text for a detailed discussion.}\label{plt2d:eLeRconstr}
\end{figure}

On the other hand, taking both $\epsilon^{e,L}$ and $\epsilon^{e,R}$ into account, we obtain simultaneous constraints on $\epsilon^{e,L}$ and $\epsilon^{e,R}$ as shown in figure\,\ref{plt2d:eLeRconstr}, where the orange and the purple regions are still allowed by Planck and CMB-S4 respectively. The permitted regions are along the diagonal region on the $\epsilon_{e,L}$-$\epsilon_{e,R}$ plane since it is where contributions from $\mathcal{O}_{1,e}^{(6)}$ and $\mathcal{O}_{2,e}^{(6)}$ cancel. This effect becomes more evident when the magnitudes of $\epsilon^{e,L}$ and $\epsilon^{e,R}$ are large, as implied by the purple regions when $|\epsilon^{e,(L,R)}|\gtrsim4$. The subfigure in the upper right corner of figure\,\ref{plt2d:eLeRconstr} is the enlarged allowed region from CMB-S4 near the origin. Since we assume neutrino flavor independence and $N_{\rm eff}$ is more sensitive to light degrees of freedom, our constraints are slightly less stringent than, but again very comparable to, those discussed in last paragraph. Our results presented in this work complement those from previous studies on NC neutrino NSIs from collider, neutrino coherent scattering and neutrino oscillation experiments.

\section{Conclusions}\label{sec:con}
Null observation of any new resonances after the discovery of the Higgs particle at the LHC has gradually changed our strategy in searching for new physics from specific UV models to model-independent studies. EFTs provide a systematic and model independent approach to heavy new physics. In the early Universe where the active fields are neutrinos, electrons, positrons and photons, the system can be described by the LEFT, even with the introduction of some new physics above the $\sim\mathcal{O}(100\rm\,MeV)$ scale. NC NSIs induced by the new physics would affect neutrino decoupling in the early Universe, thus would also modify the prediction of $N_{\rm eff}$. In light of the very precision measurements of $N_{\rm eff}$ from current Planck data and the precision target from CMB-S4, we present constraints on NC NSIs from $N_{\rm eff}$ up to dimension-7 in this work by assuming that all NC NSIs are induced by heavy mediators above $\sim\mathcal{O}(\rm 100\,MeV)$.

To that end, we adopt the strategy developed in Refs.\,\cite{Escudero:2018mvt,Escudero:2020dfa}, which permits a fast and precision calculation of $N_{\rm eff}$, and can also be easily generalized to include various new physics. The fast calculation of $N_{\rm eff}$ largely seeds in the pre-calculated collision term integrals, which are only obtained for several specific processes in the SM. In this work, we provide a complete, generic and analytical dictionary for these collision term integrals in section\,\ref{sec:SMEFT}. With our results, as long as the invariant amplitudes are known, one can refer to this dictionary to write down the Boltzmann equations, and then solve the prediction of $N_{\rm eff}$ from the SM or some new physics with few efforts. We also show an example for the application of this dictionary at the end of section\,\ref{sec:SMEFT}.

Including the NC NSIs and using the dictionary described above, we study constraints on these operators from precision measurements of $N_{\rm eff}$. Our results are presented in figure\,\ref{plt:ConstraintsNP} and summarized in table\,\ref{tab:LamLowerBound}, where the lower bounds on the scale of new physics $\Lambda$ is obtained by fixing the Wilson coefficients at unity and considering only one non-vanishing NSI operator at a time. We find that, the dimension-6 NSI operators $\mathcal{O}_{1,e}^{(6)}$ and $\mathcal{O}_{2,e}^{(6)}$ are constrained to be above $\sim331$\,GeV and $\sim240$\,GeV respectively from CMB-S4. On the other hand, due to suppression from the new physics scale, the couplings and $m_e$, dimension-7 operators $\mathcal{O}_{(5,6,7,8,9,10,11),e}^{(7)}$ only have visible corrections to $N_{\rm eff}$ when the new physics is relatively light, thus the current lower bounds on these operators are about $6$\,GeV and $3$\,GeV for $\mathcal{O}_{(7,8,9,10,11),e}^{(7)}$ and $\mathcal{O}_{(5,6),e}^{(7)}$, respectively. Operators $\mathcal{O}_{3,4,5}^{(6)}$ are not constrained from $N_{\rm eff}$ due to (1) negligible corrections to $N_{\rm eff}$ when $\Lambda\gtrsim\Lambda_W$ and (2) realization of thermal equilibrium among the three flavor neutrinos that results in vanishing contributions to $N_{\rm eff}$. Operators $\mathcal{O}_{(1,2),e}^{(7)}$ are also not constrained from $N_{\rm eff}$ due to suppression of tiny couplings.

On the other hand, we also study constraints on the Wilson coefficients with $\Lambda$ fixed at 1\,TeV and 100\,GeV. The results are shown in figure\,\ref{plt:DelNeffPlots2} by taking only one non-vanishing NSI operator into account at a time. We find that only $C_{1,e}^{(6)}$ and $C_{2,e}^{(6)}$ are constrained by $N_{\rm eff}$ since the dimension-7 operators are all suppressed by one more power of $\Lambda$, as well as $m_e$ and the small couplings. At $\Lambda=100$\,GeV, we find the magnitude of $C_{1,e}^{(6)}$ is constrained to be around 0.3 (0.1) from Planck (CMB-S4), while it is about 1.4 (0.2) for $C_{2,e}^{(6)}$ from Planck (CMB-S4). The results are summarized in eqs.\,(\ref{C61WilCoeff}-\ref{C62WilCoeff2}).

Constraints on the dimension-6 neutrino-electron NC NSI operators $\mathcal{O}_{(1,2),e}^{(6)}$ from precision measurements of $N_{\rm eff}$ are also compared with previous results from, for example, a global fitting of neutrino oscillation experiments and collider experiments. To that end, we first obtain constraints on the NC NISs using the $\epsilon$ parameterization, and then present the results in figure\,\ref{plt:eLeRconstr} and table\,\ref{tab:epsbounds} for one non-vanishing NC NSI operator at a time, and figure\,\ref{plt2d:eLeRconstr} for the inclusion of both operators. We find that constraints from precision measurements of $N_{\rm eff}$ from Planck are in general weaker than those from other experiments mentioned above. However, the improved results from CMB-S4 in future would become comparable for certain operators. Our work complements previous studies on NC NSIs from other experiments. In the future, if the cosmic neutrino background (C$\nu$B) could be directly measured, $N_{\rm eff}$ would be determined with a much better precision, and one could then expect also much stronger constraints on these NC neutrino NSIs from C$\nu$B.

\acknowledgments{We thank Shu-Yuan Guo for his valuable contribution at the early stage of this project, Miguel Escudero for helpful discussion, and the HPC Cluster of ITP-CAS for the computation support. YD and JHY were supported by the National Science Foundation of China (NSFC) under Grants No. 12022514 and No. 11875003. JHY was also supported by the National Science Foundation of China (NSFC) under Grants No. 12047503 and National Key Research and Development Program of China Grant No. 2020YFC2201501.}

\end{document}